\documentclass[runningheads]{llncs}

\usepackage{etex} % load this package right after \documentclass{} if you encounter an error saying '! No room for a new \dimen'

\usepackage{bm}
\usepackage{amsmath,amsfonts,amssymb}
\usepackage{stmaryrd}
\usepackage{times}
\usepackage[font=scriptsize,labelfont=bf]{caption}
\usepackage[inline]{enumitem}
\usepackage[ruled,vlined,linesnumbered]{algorithm2e}
\SetAlFnt{\scriptsize\sffamily} %scriptsizing fonts in algorithms
\SetAlCapNameFnt{\scriptsize\sffamily}
\SetAlCapFnt{\scriptsize\sffamily}
\usepackage[usenames,dvipsnames]{color}

\usepackage{pdfpages} % to include pdf file in the current doc, should be loaded after the color package

\usepackage[title]{appendix}

\RequirePackage[figuresleft]{rotating} %for sidewaystables
\usepackage{pdflscape}
\usepackage{booktabs}
\usepackage{array}
\usepackage{multirow}
\newcolumntype{P}[1]{>{\centering\arraybackslash}p{#1}}

\usepackage{pgf}
\usepackage{tikz}
\usepackage{tikz-qtree,tikz-qtree-compat}
\usetikzlibrary{arrows,shapes,snakes,automata,backgrounds,petri,calc,hobby,positioning,fadings,through,arrows.meta}

\usepackage{xsavebox} %for saving a tikzpicture in an xlrbox and scale it when using; seems conflicting with the bigfoot package

\usepackage{scalerel} %for scaling sub-/supscripts in math

\usepackage{graphicx}
\usepackage{subfig}
	\graphicspath{{./figures/}}
	\DeclareGraphicsExtensions{.pdf}

\usepackage{picinpar} % for using wrapfigure in theorem-like environments, e.g. example.

\usepackage{cite}

\usepackage{marvosym}

\usepackage[normalem]{ulem}  %for strikingthrough text
\newcommand\redsout{\bgroup\markoverwith{\textcolor{red}{\rule[0.5ex]{2pt}{0.4pt}}}\ULon}  %strikingthrough in red

\usepackage[hyperfootnotes=false,colorlinks,linkcolor=RedViolet,anchorcolor=black,citecolor=blue,urlcolor=black]{hyperref} % tag the hyperref with pure colors

\usepackage{etoolbox} % for overriding the paragraph style from 'it' to 'bf'
\patchcmd{\paragraph}{\itshape}{\bfseries\boldmath}{}{} % for overriding the paragraph style from 'it' to 'bf'

\let\llncssubparagraph\subparagraph
\let\subparagraph\paragraph
\usepackage{titlesec} %re-spacing around section titles: \titlespacing*{<command>}{<left>}{<before-sep>}{<after-sep>}
\makeatletter % curing a 'numbering-lost'-bug in old versions of titlesec 
\patchcmd{\ttlh@hang}{\parindent\z@}{\parindent\z@\leavevmode}{}{}
\patchcmd{\ttlh@hang}{\noindent}{}{}{}
\makeatother
\let\subparagraph\llncssubparagraph

\titlespacing*{\section}{0pt}{.9\baselineskip}{.6\baselineskip}
\titlespacing*{\subsection}{0pt}{.6\baselineskip}{.4\baselineskip}
\titlespacing*{\subsubsection}{0pt}{.3\baselineskip}{.2\baselineskip}

\input{def}

\begin{document}

\title{NIL: Learning Nonlinear Interpolants\thanks{This work has been supported through grants by NSFC under grant No.\ 61625206 and 61732001, by the CAS Pioneer Hundred Talents Program under grant No. Y9RC585036, and by the National Science Foundation Award DMS-1217054.}
}

\titlerunning{NIL: Learning Nonlinear Interpolants}

%\oomit{ % Omitted due to blind reviews.
\author{%\small
Mingshuai Chen\inst{1,2}$^{\text{(\Letter)}}$ %\orcidID{0000-0001-9663-7441}
\and Jian Wang\inst{1,2} %\orcidID{0000-0002-8840-5605}
\and Jie An\inst{3} %\orcidID{0000-0001-9260-9697}
\and Bohua Zhan\inst{1,2}$^{\text{(\Letter)}}$ %\orcidID{0000-0001-5377-9351}
\and\\ Deepak Kapur\inst{4} %\orcidID{0000-0003-2464-2895}
\and Naijun Zhan\inst{1,2}$^{\text{(\Letter)}}$ %\orcidID{0000-0003-3298-3817}
}
%\authorrunning{M.\ Chen, J.\ Wang, J.\ An, B. Zhan, N. Zhan, and D.\ Kapur}
\authorrunning{M.\ Chen et al.}
\institute{
State Key Lab. of Computer Science, Institute of Software, CAS, Beijing, China\\
\and
University of Chinese Academy of Sciences, Beijing, China\\
%\email{chenms@ios.ac.cn}
\email{\{chenms,bzhan,znj\}@ios.ac.cn}
\and
School of Software Engineering, Tongji University, Shanghai, China\\
%\email{1510796@tongji.edu.cn}
\and
Department of Computer Science, University of New Mexico, Albuquerque, USA\\
%\email{kapur@cs.unm.edu}
}
%}

\maketitle

\vspace*{-.3cm}

\setcounter{footnote}{0}

\setlength{\floatsep}{1\baselineskip}
\setlength{\textfloatsep}{1\baselineskip}
\setlength{\intextsep}{1\baselineskip}

\begin{abstract}
Nonlinear interpolants have been shown useful for the verification of programs and hybrid systems in contexts of theorem proving, model checking, abstract interpretation, etc. The underlying synthesis problem, however, is challenging and existing methods have limitations on the form of formulae to be interpolated. We leverage classification techniques with space transformations and kernel tricks as established in the realm of machine learning, and present a counterexample-guided method named NIL for synthesizing polynomial interpolants, thereby yielding a unified framework tackling the interpolation problem for the general quantifier-free theory of nonlinear arithmetic, possibly involving transcendental functions. We prove the soundness of NIL and propose sufficient conditions under which NIL is guaranteed to converge, i.e., the derived sequence of candidate interpolants converges to an actual interpolant, and is complete, namely the algorithm terminates by producing an interpolant if there exists one. The applicability and effectiveness of our technique are demonstrated experimentally on a collection of representative benchmarks from the literature, where in particular, our method suffices to address more interpolation tasks, including those with perturbations in parameters, and in many cases synthesizes simpler interpolants compared with existing approaches. 
\end{abstract}

\section{Introduction}\label{sec_intro}

Interpolation-based technique provides a powerful mechanism for local and modular reasoning, thereby improving scalability of various verification techniques, e.g., theorem proving, model checking and abstract interpretation, to name just a few. The study of interpolation was pioneered by Kraj{\'{\i}{\v c}}ek~\cite{krajicek97} and Pudl\'{a}k~\cite{pudlak97} in connection with theorem proving, by McMillan~\cite{mcmillan03} in the context of model checking, by Graf and Sa\"{i}di~\cite{GS97}, McMillan~\cite{mcmillan04} and Henzinger et al.~\cite{HJMM04} pertaining to abstraction like CEGAR~\cite{ClarkeGJLV00}, and by Wang et al.~\cite{wang11} in the context of learning-based invariant generation. Developing efficient algorithms for generating interpolants for various theories and their combination has become an active research area, see e.g.,~\cite{mcmillan04,YM05,KMZ06,RS10,KV09,CGS08,mcmillan08}.
%We refer the readers further to~\cite{SPWK10} for an ordered family of interpolation systems due to the logical strength of the synthesized interpolants.

Though established methods addressing interpolant generation for Presburger arithmetic, decidable fragments of first-order logic, theory of equality over uninterpreted functions (EUFs) as well as their combination have been extensively studied in the literature, there appears to be little work on synthesizing nonlinear interpolants.
%, albeit the fact that nonlinear polynomial constraints have been shown useful to express invariant properties for software involving number-theoretic functions as well as hybrid systems, cf., e.g.,~\cite{ZZKL12,ZZK13}.
Dai et al. proposed an algorithm in~\cite{DXZ13} for generating interpolants for nonlinear polynomial inequalities based on the existence of a witness guaranteed by Stengle's Positivstellensatz~\cite{Stengle} that can be computed using semi-definite programming (SDP). A major limitation of this method is that the two mutually contradictory formulas to be interpolated must share the same set of variables.
Okudono et al. extended~\cite{DXZ13} in~\cite{DBLP:conf/aplas/OkudonoNKSKH17} to cater for the so-called sharper and simpler interpolants by developing a continuous fraction-based algorithm that rounds off numerical solutions.
In~\cite{GDX16}, Gan et al. considered the interpolation for inequalities combined with EUFs by employing the hierarchical calculus framework proposed in \cite{DBLP:conf/cade/Sofronie-Stokkermans06} (and its extension~\cite{Sofronie-Stokkermans16}), while the inequalities are limited to be of the concave quadratic form.
In~\cite{DBLP:conf/tacas/GaoZ16}, Gao and Zufferey transformed proof traces from $\delta$-complete decision procedures into interpolants, composed of Boolean combinations of linear constraints, which can deal with certain transcendental functions beyond polynomials. The techniques of encoding interpolants as logical combinations of linear constraints, including~\cite{DBLP:conf/tacas/GaoZ16},~\cite{KupferschmidB11} and~\cite{Sharma12}, however, yield potentially large interpolants (requiring even an infinite length in the worst case) and their usage thus becomes difficult in practical applications (cf. Example~\ref{exmp:tacas16}).

Interpolants can be viewed as classifiers that distinguish, in the context of program verification for instance, positive program states from negative ones (unreachable/error states) and consequently the state-of-the-art classification algorithms can be leveraged for synthesizing interpolants. The universal applicability of classification techniques substantially extends the scope of theories admitting interpolant generation. This idea was first employed by Sharma et al. in~\cite{Sharma12}, which infers linear interpolants through hyperplane-classifiers generated by support vector machines (SVMs)~\cite{VapLer63,Boser92} whilst handles superficial nonlinearities by assembling interpolants in the form purely of conjunctions (or dually, disjunctions) of linear half-spaces, which addresses only a limited category of formulae featuring nonlinearities. The learning-based paradigm has also been exploited in the context of nonlinear constraint solving, see e.g.,~\cite{DBLP:conf/ijcai/DathathriAGM17}.

%\paragraph*{\it Contributions and Outline.}
In this paper, we present a classification-based learning method for the synthesis of polynomial interpolants for the quantifier-free theory of nonlinear arithmetic. Our approach is based on techniques of space transformations and kernel tricks pertinent to SVMs that have been well-developed in the realm of machine learning. Our method is described by an algorithm called NIL (and its several variants) that adopts the counterexample-guided inductive synthesis framework~\cite{JhaGST10,Solar-LezamaRBE05}. We prove the soundness of NIL and propose sufficient conditions under which NIL is guaranteed to converge, that is, the derived sequence of classifiers (candidate interpolants) converges to an actual interpolant, and is complete, i.e., if an interpolant exists, the method terminates with an actual interpolant. In contrast to related work on generation of nonlinear interpolants, which restrict the input formulae, our technique provides a uniform framework, tackling the interpolation problem for the general quantifier-free theory of nonlinear arithmetic, possibly involving transcendental functions. The applicability and effectiveness of NIL are demonstrated experimentally on a collection of representative benchmarks from the literature; as is evident from experimental results, our method is able to address more demands on the nature of interpolants, including those with perturbations in parameters (due to the robustness inherited from SVMs); in many cases, it synthesizes simpler interpolants compared with other approaches, as shown by the following example.

\begin{example}[\cite{DBLP:conf/tacas/GaoZ16}]\label{exmp:tacas16}
Consider two mutually contradictory inequalities $\phi \define y\ge x^2$ and $\psi \define y\le -\cos(x) + 0.8$. Our NIL algorithm constructs a single polynomial inequality $I \define 15 x^2 < 4 + 20y$ as the interpolant, namely, $\phi \models I$ and $I \wedge \psi$ is unsatisfiable; while the interpolant generated by the approach in~\cite{DBLP:conf/tacas/GaoZ16}, only when provided with sufficiently large finite domains, e.g., $x\in [-\pi, \pi]$ and $y\in [-0.2, \pi^2]$, is $y>1.8\lor(0.59\leq y\leq 1.8\land -1.35\leq x\leq 1.35)\lor (0.09\leq y<0.59\land -0.77\leq x\leq0.77)\lor (y\geq 0\land -0.3\leq x\leq 0.3)$. As will be discussed later, we do not need to provide a priori information to our algorithm such as bounds on variables.
\end{example}

The rest of the paper is organized as follows. Sect.~\ref{sec_preliminaries} introduces some preliminaries on Craig interpolants and SVMs. In Sect.~\ref{sec_learning}, we present the NIL algorithm dedicated to synthesizing nonlinear interpolants, followed by the analysis of its soundness, conditional completeness and convergence in Sect.~\ref{sec_theories}. Sect.~\ref{sec_experiments} reports several implementation issues and experimental results on a collection of benchmarks (with the robustness discussed in Sect.~\ref{subsec_robustness}). The paper is then concluded in Sect.~\ref{sec_conclusion}.

\oomit{ % merged previously.
\paragraph*{\it Related Work.}
%We place our work in the context of existing work on the interpolation over nonlinear arithmetic.
Dai et al. proposed an algorithm in~\cite{DXZ13} for generating interpolants for nonlinear polynomial inequalities based on the existence of a witness guaranteed by Stengle's Positivstellensatz~\cite{Stengle} that can be computed using semi-definite programming (SDP). A major limitation of this method is that the two mutually contradictory formulas to be interpolated must share the same set of variables.
Okudono et al. extended~\cite{DXZ13} in~\cite{DBLP:conf/aplas/OkudonoNKSKH17} to cater for the so-called sharper and simpler interpolants by developing a continuous fraction-based algorithm that rounds off numerical solutions.
In~\cite{GDX16}, Gan et al. considered the interpolation for inequalities combined with EUFs by employing the hierarchical calculus framework proposed in \cite{DBLP:conf/cade/Sofronie-Stokkermans06}, while the inequalities are limited to be of the concave quadratic form.

In~\cite{DBLP:conf/tacas/GaoZ16}, Gao and Zufferey transformed proof traces from $\delta$-complete decision procedures into interpolants composing of Boolean combinations of linear constraints, which can deal with certain transcendental functions beyond polynomials. The techniques of encoding interpolants as logical combinations of linear constraints, including~\cite{DBLP:conf/tacas/GaoZ16} and~\cite{Sharma12}, however, yield potentially large interpolants (even of infinite length in the worst case) and their usage thus becomes difficult in practical applications. Our method instead constructs a single polynomial inequality as an interpolant, for example, given two mutually contradictory inequalities $\phi \define y\ge x^2$ and $\psi \define y\le -\cos(x) + 0.8$, the interpolant generated by our NIL algorithm is $15 x^2 < 4 + 20y$, while the interpolant generated by the approach in~\cite{DBLP:conf/tacas/GaoZ16}, only when provided with sufficiently large finite domains, e.g., $x\in [-\pi, \pi]$ and $y\in [-0.2, \pi^2]$, is $y>1.8\lor(0.59\leq y\leq 1.8\land -1.35\leq x\leq 1.35)\lor (0.09\leq y<0.59\land -0.77\leq x\leq0.77)\lor (y\geq 0\land -0.3\leq x\leq 0.3)$.
%\znjSide{For infinite domain, \cite{DBLP:conf/tacas/GaoZ16} still works?}
%\chenSide{Not really. It works for finite domains only.}
}

\section{Preliminaries}\label{sec_preliminaries}

Let $\NN$, $\QQ$ and $\RR$ be the set of natural, rational and real numbers, respectively. We denote by $\RR[\xx]$ the polynomial ring over $\RR$ with variables $\xx=(\xx_1,\ldots,\xx_n)$, and $\|\xx\|$ denotes the $\ell^2$-norm~\cite{bourbaki1987topological}. For a set $X \subseteq \RR^n$, its convex hull is denoted by $\conv{X}$. For $\vec{x}, \vec{x}' \in X$, $\dist{\vec{x},\vec{x}'} = \|\vec{x} - \vec{x}'\|$ denotes the Euclidean distance between two points, which generalizes to $\dist{\vec{x},X'} = \min_{\vec{x}' \in X'} \dist{\vec{x},\vec{x}'}$. Given $\delta \ge 0$, define $\mathcal{B}(\vec{x},\delta) = \{\vec{x}' \in \RR^n | \|\vec{x}'-\vec{x}\| \le \delta\}$ as the closed ball of radius $\delta$ centered at $\vec{x}$. Consider the quantifier-free fragment of a first-order theory of polynomials over the reals, denoted by $\PT$, in which a formula $\varphi$ is of the form
\begin{equation*}
\varphi \define p(\xx) \diamond 0 \ \mid\  \varphi \wedge \varphi \ \mid \ \varphi \vee \varphi \ \mid \ \neg \varphi
\end{equation*}
where $p(\xx) \in \RR[\xx]$ and $\diamond \in \{<, >, \le, \ge, =\}$. A natural extension of our method to cater for more general nonlinearities involving transcendental functions will be demonstrated in subsequent sections. In the sequel, we use $\bot$ to stand for \emph{false} and $\top$ for \emph{true}. Let $\RR[\xx]_m$ consist of all polynomials $p(\xx)$ of degree $\le m \in \NN$. We abuse the notation $\varphi \in \RR[\xx]_m$ to abbreviate $\varphi \define p(\xx) \diamond 0$ and $p(\xx) \in \RR[\xx]_m$ if no ambiguity arises.
%A polynomial $p(\xx)$ is said to be in $\RR[\xx]^m$ iff $\deg p(\xx) = m \in \NN$.

Given formulas $\phi$ and $\psi$ in a theory $\TT$, $\phi$ is \emph{valid} w.r.t. $\TT$, written as $\models_{\TT} \phi$,  iff $\phi$ is true in all models of $\TT$; $\phi$ \emph{entails} $\psi$ w.r.t. $\TT$, written as $\phi \models_{\TT} \psi$, iff every model of $\TT$ that makes $\phi$ true makes $\psi$ also true; $\phi$ is \emph{satisfiable} w.r.t. $\TT$, iff there is a model of $\TT$ in which $\phi$ is true; otherwise \emph{unsatisfiable}. It follows that $\phi$ is unsatisfiable iff $\phi \models_{\TT} \bot$. The set of all the models that make $\phi$ true is denoted by $\llbracket \phi \rrbracket_{\TT}$.

\subsection{Craig Interpolant}\label{sub_sec_craig}

Craig showed in \cite{craig_1957} that given two formulas $\phi$ and $\psi$ in a first-order logic $\TT$ s.t. $\phi \models_{\TT} \psi$, there always exists an \emph{interpolant} $I$ over the common symbols of $\phi$ and $\psi$ s.t. $\phi \models_{\TT} I$ and $I \models_{\TT} \psi$. In the verification literature, this terminology has been abused by~\cite{mcmillan04}, which defined an interpolant over the common symbols of $\phi$ and $\psi$ as
%where a \emph{reverse interpolant} (coined by Kov\'{a}cs and Voronkov in \cite{KV09}) $I$ over the common symbols of $\phi$ and $\psi$ is defined by 

\begin{definition}[Interpolant] \label{def_interpolant}
  Given $\phi$ and $\psi$ in a theory $\TT$ s.t.
  $\phi \wedge \psi \models_{\TT} \bot$, a formula $I$ is
  a \emph{(reverse) interpolant} of $\phi$ and
  $\psi$ if
  %\begin{enumerate}
  \emph{(i)} $\phi \models_{\TT} I$;
  \emph{(ii)} $I \wedge \psi \models_{\TT} \bot$; and
  \emph{(iii)}  $I$ contains only common symbols shared by $\phi$ and
  $\psi$.
 % \end{enumerate}
\end{definition}

It is immediately obvious that $\phi \models_{\TT} \psi$ iff $\phi \wedge \neg \psi \models_{\TT} \bot$, namely, $I$ is an interpolant of $\phi$ and $\psi$ iff $I$ is a reverse interpolant in McMillan's sense of $\phi$ and $\neg \psi$. We follow McMillan in continuing to abuse the terminology.
%is a reverse interpolant of $\phi$ and $\neg \psi$. We abuse the terminology in the sequel by calling reverse interpolants as interpolants.

\subsection{Support Vector Machines}\label{sec_svms}

In machine learning, support vector machines~\cite{VapLer63,Boser92} are supervised learning models for effective classification based on convex optimization. In a binary setting, we are given a training dataset $X = X^+ \uplus X^-$ of $n$ sample points $\{(\vec{x}_1,y_1),(\vec{x}_2,y_2),\ldots,(\vec{x}_n,y_n)\}$, where $\vec{x}_i \in \RR^d$, and $y_i$ is either 1, indicating a positive sample $\vec{x}_i \in X^+$, or -1, indicating a negative one in $X^-$. The goal of classification here is to find a potential hyperplane (a.k.a. \emph{linear classifier}) to separate the positive samples from the negative ones. There however might be various or even infinite number of separating hyperplanes, and an SVM aims to construct a separating hyperplane that yields the largest distance (so-called \emph{functional margin}) to the nearest positive and negative samples. Such a classification hyperplane is called the \emph{optimal-margin classifier} while the samples closest to it are called the \emph{support vectors}.

\subsubsection{Linear SVMs.}

Assume that $X^+$ and $X^-$ are \emph{linearly separable}, meaning that there exists a \emph{linear separating hyperplane} $\vec{w}^\mathrm{T} \vec{\xx} + b = 0$ such that
$y_i (\vec{w}^\mathrm{T} \vec{x}_i + b) > 0$, for all $(\vec{x}_i,y_i) \in X.$
Then the functional margin can be formulated as
\begin{equation*}
\gamma \define 2 \min\nolimits_{1 \le i \le n} 1/{\| \vec{w} \|} \lvert \vec{w}^\mathrm{T} \vec{x}_i + b \rvert .
\end{equation*}
Linear SVMs are committed to finding appropriate parameters $(\vec{w}, b)$ that maximize the functional margin while adhering to the constraints of separability,
%\begin{eqnarray*}
%\underset{\vec{w}, b}{\text{maximize}} & & \min\limits_{1 \le i \le n} \frac{1}{\| \vec{w} \|} \lvert \vec{w}^\mathrm{T} \vec{x}_i + b \rvert \\
%\text{subject to} & & y_i (\vec{w}^\mathrm{T} \vec{x}_i + b) > 0, \ \ i = 1, 2, \ldots, n,
%\end{eqnarray*}
which reduces equivalently to the following convex quadratic optimization problem \cite{Bishop06} that can be efficiently solved by off-the-shelf packages for quadratic programming:
\begin{equation}\label{eq_quadratic}
\underset{\vec{w}, b}{\text{minimize}} \ \ \frac{1}{2} \vec{w}^\mathrm{T} \vec{w} \quad
\text{subject to} \ \ y_i (\vec{w}^\mathrm{T} \vec{x}_i + b) \ge 1, \ \ i = 1, 2, \ldots, n.
\end{equation}
%\begin{equation}
%\begin{split}
%\underset{\vec{w}, b}{\text{minimize}} \ \ &\frac{1}{2} \vec{w}^\mathrm{T} \vec{w} \\
%\text{subject to} \ \ &y_i (\vec{w}^\mathrm{T} \vec{x}_i + b) \ge 1, \ \ i = 1, 2, \ldots, n.
%\end{split}\label{eq_quadratic}
%\end{equation}

\begin{lemma}[Correctness of SVMs \cite{Sharma12}]
Given positive samples $X^+$ which are linearly separable from negative samples $X^-$, SVMs produce, under computations of infinite precision, a half-space $h$ s.t. $\forall \vec{x} \in X^+.\ h(\vec{x})>0$ and $\forall \vec{x} \in X^-.\ h(\vec{x})<0$.
\label{lemma_SVMs_corr}
\end{lemma}

\begin{corollary}[Separation of Convex Hulls \cite{DBLP:conf/icml/BennettB00}]
The half-space $h$ in Lemma~\ref{lemma_SVMs_corr} satisfies that $\forall \vec{x} \in \conv{X^+}.\ h(\vec{x})>0$ and $\forall \vec{x} \in \conv{X^-}.\ h(\vec{x})<0$.
\label{corollary_SVMs_conv}
\end{corollary}

\tikzset{
	>=latex,
	leftNode/.style={circle,minimum width=.3ex,scale=0.5,fill=none,draw},
	rightNode/.style={circle,minimum width=.3ex,scale=0.5,fill=black,draw}
}
\newcommand{\edge}{2}
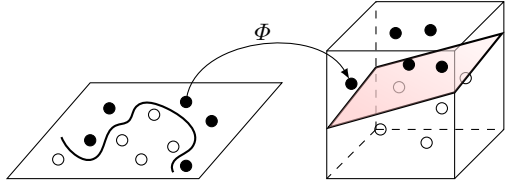
\begin{figure}[t]
	\centering
	\begin{tikzpicture}[
	scale=.85,
	classifier/.style={thick},
	every node/.style={color=black},
	]
	
	\draw[-]
	(0,0) -- (3,0) -- (4.3,1.5) -- (1.3,1.5) -- cycle;
	
	\foreach \Point in {(.8,0.3), (1.8,0.6), (2.3,1), (2.6,0.5), (2.1,0.3)}{
		\draw \Point node[leftNode]{};
	}
	
	\foreach \Point in {(1.3,0.6), (1.6,1.1), (2.8,1.2), (3.2,0.9), (2.8,0.2)}{
		\draw \Point node[rightNode]{};
	}
	
	\draw[classifier] (.85,.65) to[curve through={(1.1,.35) .. (1.33,.3) .. (1.7,.83) .. (1.9,.9) .. (2.0, 1.1) .. (2.5,1.1) .. (2.9,.5) .. (2.58,.2)}] (2.6,.1);
	
	\coordinate (O) at (5,0,0);
	\coordinate (A) at (5,\edge,0);
	\coordinate (B) at (5,\edge,-\edge);
	\coordinate (C) at (5,0,-\edge);
	\coordinate (D) at (5+\edge,0,0);
	\coordinate (E) at (5+\edge,\edge,0);
	\coordinate (F) at (5+\edge,\edge,-\edge);
	\coordinate (G) at (5+\edge,0,-\edge);
	
	\draw (O) -- (D) -- (E) -- (A) -- cycle;
	\draw (D) -- (G) -- (F) -- (B) -- (A) (E) --(F);
	\path[dashed] (C) edge (O) edge (G) edge (B);
	%% Following is for debugging purposes so you can see where the points are
	%% These are last so that they show up on top
	%\foreach \xy in {O, A, B, C, D, E, F, G}{
	%    \node at (\xy) {\xy};
	%}
	
	\foreach \Point in {(5.07,0,-2),(5.9,1.2,-.6),(7,1.4,-.45),(6.3,.6,-1.3),(6.1,.1,-1.2)}{
		\draw \Point node[leftNode]{};
	}
	\foreach \Point in {(5,1.1,-1),(5.7,1.9,-.98),(6.2,1.9,-1.1)}{
		\draw \Point node[rightNode]{};
	}
	
	\draw[classifier,fill=pink!30!white, postaction={path fading=north, fading angle=-45, fill=pink},fill opacity=0.5] (5.03,.8,0) -- (7,1.35,0) -- (6.97,1.5,-2) -- (5,.97,-2) -- cycle;
	\foreach \Point in {(5.9,1.4,-1),(6.3,1.25,-1.3)}{
		\draw \Point node[rightNode]{};
	}
	
	\draw[->] (2.8,1.2,0) to[out=85,in=120] (4.97,1.17,-1);
	\node at (4,2.3,0) {$\Phi$};
	
	\end{tikzpicture}
	\caption{Mapping from a two-dimensional input space into a three-dimensional feature space with linear separation thereof.}
	\label{fig_svm_nonlinear}
\end{figure}

\subsubsection{Nonlinear SVMs.}

When $\phi$ and $\psi$ are formulas over nonlinear arithmetic, often after sampling $X$, it is not possible to find a linearly separable hyperplane in the common variables. However, a nonlinear surface that can be described as a linear hyperplane in the space of monomials of bounded degree may separate $X^+$ and $X^-$. The above construction is generalized by introducing a transformation from $\mathbb{R}^d$ to $\mathbb{R}^{\tilde{d}}$, the vector space of monomials in the common variables up to some bounded degree, with $y_i (\vec{w}^\mathrm{T} \vec{x}_i + b) \ge 1$ in~\eqref{eq_quadratic} replaced by
$y_i (\vec{w}^\mathrm{T} \Phi(\vec{x}_i) + b) \ge 1$, where $\Phi$ is a linear expression in monomials in the common variables up to a bounded degree. Here, the vectors $\Phi(\xx)$ span the \emph{feature space}.
\oomit{
Putting aside numerical errors involved in the computation that would be addressed later, the correctness of SVMs stated in Lemma \ref{lemma_SVMs_corr} follows from the fact that $X^+$ and $X^-$ are linearly separable, this however is often violated when $X$ is sampled from constraints over nonlinear arithmetic. A well-developed method to resolve this issue is by introducing a nonlinear \emph{transformation} $\Phi\colon \mathbb{R}^d \mapsto \mathbb{R}^{\tilde{d}}$, which maps a vector $\vec{x}\in\mathbb{R}^d$ in the original space to $\Phi(\vec{x})\in\mathbb{R}^{\tilde{d}}$ in the \emph{feature space}. Suppose that there exists a nonlinear separating hyperplane $\vec{w}^\mathrm{T}\Phi(\xx)+b=0$ such that
$y_i (\vec{w}^\mathrm{T} \Phi(\vec{x}_i) + b) \ge 1$, $\forall (\vec{x}_i,y_i) \in X$.
Then the optimization problem~\eqref{eq_quadratic} becomes:
\begin{equation}
\begin{split}
\underset{\vec{w}, b}{\text{minimize}} \ \ &\frac{1}{2} \vec{w}^\mathrm{T} \vec{w} \\
\text{subject to} \ \ &y_i (\vec{w}^\mathrm{T} \Phi(\vec{x}_i) + b) \ge 1, \ \ i = 1, 2, \ldots, n.
\end{split}\label{eq_transformation}
\end{equation}
However, computing the map $\Phi(\vec{x})$ would be rather computationally expensive in case of a large dimension $\tilde{d}$ of the feature space, where the so-called \emph{kernel trick} \cite{Boser92} is introduced to compute the separating hyperplane without explicitly carrying out the map into the feature space.
}

Consider the Lagrangian dual~\cite{Boser92} of the modified optimization problem:
%~\eqref{eq_transformation}:
\begin{eqnarray*}
\underset{\alpha}{\text{minimize}} & & \frac{1}{2}\sum\nolimits_{i=1}^{n}\sum\nolimits_{j=1}^{n}\alpha_i\alpha_j y_i y_j \Phi(\vec{x}_i)^\mathrm{T}\Phi(\vec{x}_j)-\sum\nolimits_{i=1}^{n}\alpha_i
\\
\text{subject to} & & \sum\nolimits_{i=1}^{n}\alpha_i y_i = 0, \ \text{and} \ \alpha_i \ge 0 \ \text{for} \ i = 1, 2, \ldots, n.
\end{eqnarray*}
A \emph{kernel function} $\kernel\colon \mathbb{R}^d \times \mathbb{R}^d \mapsto \mathbb{R}$ is defined as $\kernel(\xx, \xx') \define \Phi(\xx)^\mathrm{T}\Phi(\xx')$. The introduction of the dual problem and the kernel function~\cite{Boser92} reduces the computational complexity essentially from $\mathcal{O}(\tilde{d})$ down to $\mathcal{O}(d)$. For the sake of post-verifying a candidate interpolant given by SVMs, we adopt an inhomogeneous polynomial kernel function of the form
\begin{equation*}
\kernel(\xx, \xx') \define (\beta \xx^\mathrm{T} \xx'+ \theta)^m,
\end{equation*}
where $m$ is the polynomial degree describing complexity of the feature space, $\theta \ge 0$ is a parameter trading off the influence of higher-order versus lower-order terms in the polynomial, and $\beta$ is a scalar parameter. Henceforth, the optimal-margin classifier (if there exists one) can be derived as
$
\vec{w}^\mathrm{T}\Phi(\xx)=\sum_{i=1}^{n}\alpha_i \kernel(\vec{x}_i, \xx)=0,
$
%\begin{equation*}
%\vec{w}^\mathrm{T}\Phi(\xx)=\sum\nolimits_{i=1}^{n}\alpha_i \kernel(\vec{x}_i, \xx)=0,
%\end{equation*}
with $\vec{x}_i$ being a support vector iff $\alpha_i > 0$. In practice, usually a large amount of $\alpha_i$s turn out to be zero and this leads to a simple representation of a classifier. Fig. \ref{fig_svm_nonlinear} illustrates the intuitive idea of the transformation from the original input space to the feature space. We will show in the sequel that the resulting classifier can be viewed as a candidate interpolant, while its optimal-margin feature contributes to a certain ``medium'' logical strength of the interpolant, which is thus robust to perturbations (in the feature space) in the formulae to be interpolated.

\section{Learning Interpolants}\label{sec_learning}

In this section, we present the NIL algorithm for
synthesizing nontrivial (reverse) Craig interpolants for the
quantifier-free theory of nonlinear arithmetic. It takes as input a
pair $\langle \phi, \psi \rangle$ of formulas in $\PT$ as well as a
positive integer $m$, and aims to generate an interpolant $I$ of maximum degree $m$, i.e., $I \in \RR[\xx]_m$, if it
exists, such that $\phi \models_{\PT} I$ and $I \wedge \psi
\models_{\PT} \bot$. Here, $\langle \phi, \psi \rangle$ can be decorated as $\langle \phi(\xx,\yy), \psi(\xx,\zz) \rangle$ with variables involved in the predicates, and thus $\xx$ denotes variables that are common to $\phi$ and $\psi$. %~\chenSide{common/uncommon variables clarified.}
In the sequel, we drop the subscript $\PT$ in $\models_{\PT}$ and $\llbracket \cdot \rrbracket_{\PT}$ wherever the context is unambiguous.
%
%\kapur{Block diagram goes here.}

Due to the decidability of the first-order theory of real-closed
fields established by Tarski~\cite{Tarski51}, $\PT$ admits \emph{quantifier
elimination} (QE). This means that the satisfiability of any formula in
$\PT$ can be decided (in doubly exponential time in the number of variables for the worst
case). If the formula is satisfiable, models satisfying the formula
can also be constructed algorithmically (following the same time complexity). Though the introduction of general forms of transcendental functions renders the underlying theory undecidable, there does exist certain extension of $\PT$ with transcendental functions (involving exponential functions, logarithms and trigonometric functions), e.g. that identified by Strzebo{\'n}ski in~\cite{Strzebonski11} and references therein, which still admits QE. This allows a straightforward extension of NIL to such a decidable fragment involving transcendental functions. Specifically, the decidability remains when the transcendental functions involved are real univariate exp-log functions~\cite{DBLP:conf/issac/Strzebonski08} or tame elementary functions~\cite{DBLP:conf/issac/Strzebonski09} which admit a real root isolation algorithm.
%
%\znjSide{Do you need to deal with uncommon variables?}
%\chenSide{Yes indeed. We will clarify here common/uncommon variables appearing in $\phi$ and $\psi$.}

\subsection{The Core Algorithm}

The basic idea of \ref{alg_nil} is to view interpolants as classifiers and use
SVMs with the kernel trick to perform effective classification. The
algorithm is based on the sampling-guessing-refining technique: in
each iteration, it is fed with a classifier
(candidate interpolant) for a finite set of sample points from
$\llbracket \phi \rrbracket$ and $\llbracket \psi
\rrbracket$ (line~\ref{alg_nil_svm}), and verify the candidate (line~\ref{alg_nil_checking}) by checking the entailment problem that defines an interpolant (as in Def.~\ref{def_interpolant}).
If the verification succeeds, the interpolant is returned
as the final result. Otherwise, a set of counterexamples is obtained (line~\ref{alg_nil_findinstance_pos} and~\ref{alg_nil_findinstance_neg}) as new sample points to further refine the classifier.
In what follows, we explain the steps of the interpolation procedure
in more detail.

\subsubsection{Initial sampling.}

The algorithm begins by checking the satisfiability of $\phi \wedge
\psi$. If the formula is satisfiable, it
is then impossible to find an interpolant, and the algorithm stops declaring no interpolant exists.

Next, the algorithm attempts to sample points from both $\llbracket
\phi \rrbracket$ and $\llbracket \psi \rrbracket$. This initial
sampling stage can usually be done efficiently using the Monte Carlo
method, e.g. by (uniformly) scattering a number of random points over
certain bounded range and then selecting those fall in
$\llbracket \phi \rrbracket$ and $\llbracket \psi
\rrbracket$ respectively. However, this method fails when one or both of the
predicates is very unlikely to be satisfied. One common example is
when the predicate involves equalities. For such situations, solving
the satisfiability problem using QE is guaranted to succeed in producing the sample points.
%guarantees to succeed in producing the sample points, albeit less efficient.

To meet the condition that the generated interpolant can only involve
symbols that are common to $\phi$ and $\psi$, we can project the points sampled from $\llbracket \phi \rrbracket$ (resp. $\llbracket \psi
\rrbracket$) to the space of $\xx$ by simply dropping the
components that pertain to $\yy$ (resp. $\zz$) and thereby obtain sample points in
$X^+$ (resp. $X^-$).

\begin{minipage}[c]{.62\linewidth}
	\hspace*{-.7cm}
	\SetAlgoRefName{NIL}
	\begin{algorithm}[H]
		\caption{Learning nonlinear interpolant}\label{alg_nil}
		\SetKwInOut{Input}{input}\SetKwInOut{Output}{output}\SetNoFillComment
		\Input{$\phi$ and $\psi$ in $\PT$ over common variables $\xx$;\\
			$m$, degree of the polynomial kernel, and hence\\
			$\quad\ $ maximum degree of the interpolant.
		}
		\tcc{checking unsatisfiability}
		\If{$\phi \wedge \psi \not\models \bot$\label{alg_nil_unsatisfiability}}
		{
			\tcc{no interpolant exists}
			\textbf{abort}\;\label{alg_nil_abort1}
		}
		\tcc{generating initial sample points}
		$\langle X^+, X^- \rangle \leftarrow \textbf{Sampling}(\phi, \psi)$\;\label{alg_nil_sampling}
		\tcc{counterexample-guided learning}
		\While{$\top$\label{alg_nil_loop}}
		{
			\tcc{generating a classifier by SVMs}
			$C \leftarrow \textbf{SVM}(X^+,X^-,m)$\;\label{alg_nil_svm}
			\tcc{checking classification result}
			\If{$C = \mathrm{Failed}$}
			{
				\tcc{no interpolant exists in $\RR[\xx]_m$}
				\textbf{abort}\;\label{alg_nil_abort2}
			}
			\tcc{classifier as candidate interpolant}
			\Else
			{
				$I \leftarrow C$\;
			}
			%	\tcc{rounding by rational recovery}
			%	$I \leftarrow \textbf{Rationalize}(C)$\;
			\tcc{valid interpolant found}
			\If{$\phi \models I \ \mathrm{and}\ I \wedge \psi \models \bot$\label{alg_nil_checking}}
			{
				\Return $I$\;\label{alg_nil_return}
			}
			\tcc{adding counterexamples}
			\Else
			{
				$X^+ \leftarrow X^+ \uplus \textbf{FindInstance}(\phi \wedge \neg I)$\;\label{alg_nil_findinstance_pos}
				$X^- \leftarrow X^- \uplus \textbf{FindInstance}(I \wedge \psi) $\;\label{alg_nil_findinstance_neg}
			}
		}
	\end{algorithm}
\end{minipage}
%\hfill
\begin{minipage}[c]{.33\linewidth}
	\tikzset{
		>=latex,
		leftNode/.style={circle,minimum width=.2ex,scale=0.3,fill=none,draw},
		rightNode/.style={circle,minimum width=.2ex,scale=0.3,fill=black,draw}
	}
	\centering
	\begin{tikzpicture}[
	scale=.85,
	classifier/.style={thick},
	every node/.style={color=black},
	]
	
	\draw (0,0) circle (1);
	\draw (2.4,0) circle (1);
	
	\foreach \Point in {(-.2,0), (-.1,.27), (-.4,.3), (-.5,-.1)}{
		\draw \Point node[rightNode]{};
	}
	
	\draw node[rightNode,fill=red,color=red] at (.8,-.1) {};
	
	\foreach \Point in {(1.4,0), (1.6,.2), (1.7,-.3), (1.5,-.2)}{
		\draw \Point node[leftNode]{};
	}
	
	\draw[classifier] (.8,1.2) -- (.4,-1.2);
	\draw[classifier,color=red] (.9,1.2) -- (1.3,-1.2);
	
	\node at (0,.8) {\tiny $\llbracket \phi \rrbracket$};
	\node at (2.4,.8) {\tiny $\llbracket \psi \rrbracket$};
	\end{tikzpicture}
	\captionof{figure}{In NIL, a candidate interpolant (black line as its boundary) is refined to an actual one (red line as its boundary) by adding a counterexample (red dot).}
	\label{fig_nil_core}
	\vspace*{.6cm}
	\begin{tikzpicture}[
	scale=.85,
	classifier/.style={thick},
	every node/.style={color=black},
	]
	
	\draw (0,0) circle (1);
	\draw (2,0) circle (1);
	
	\foreach \Point in {(-.2,0), (-.1,.27), (-.4,.3), (-.5,-.1)}{
		\draw \Point node[rightNode]{};
	}
	
	\draw node[rightNode,fill=red,color=red] at (.7,-.07) {};
	
	\foreach \Point in {(1,0), (1.2,.2), (1.3,-.3), (1.1,-.2)}{
		\draw \Point node[leftNode]{};
	}
	
	\draw[classifier] (.6,1.2) -- (.2,-1.2);
	\draw[classifier,color=red] (.65,1.2) -- (1.05,-1.2);
	
	\draw[{Latex[length=.5mm,width=.5mm]}-{Latex[length=.5mm,width=.5mm]}] (.41,-.02) to (.65,-.06);
	\node at (.56,.1) {\tiny $\delta$};
	
	\node at (0,.8) {\tiny $\llbracket \phi \rrbracket$};
	\node at (2,.8) {\tiny $\llbracket \psi \rrbracket$};
	\end{tikzpicture}
	\captionof{figure}{In NIL$_\delta$, a counterexample (red dot) stays at least a distance of $\delta$ away from the candidate interpolant (black line as its boundary) to be refined, leading to an interpolant (red line as its boundary) with tolerance $\delta$.}
	\label{fig_nil_delta}
\end{minipage}

\subsubsection{Entailment checking.}

The correctness of SVM given in Lemma~\ref{lemma_SVMs_corr} only
guarantees that the candidate interpolant separates the finite set of
points sampled from $\llbracket \phi \rrbracket$ and $\llbracket \psi
\rrbracket$, not necessarily the entirety of the two sets. Hence,
post-verification by checking the entailment problem (line~\ref{alg_nil_checking}) is needed for the candidate to be claimed as an
interpolant of $\phi$ and $\psi$. This can be achieved by solving the equivalent QE problems $\forall \xx.\ \phi(\xx,\yy)\vert_{\xx} \implies I(\xx)$ and $\forall \xx.\ I(\xx) \wedge \psi(\xx,\zz)\vert_{\xx} \implies \bot$, where $\cdot\vert_{\xx}$ is the \emph{projection} to the common space over $\xx$. The candidate will be returned as an actual interpolant if both formulae reduce to $\top$ after eliminating the universal quantifiers. The satisfiability checking at line~\ref{alg_nil_unsatisfiability} can be solved analogously. Granted, the entailment checking can also be encoded in SMT techniques by asking the satisfiability of the negation of the universally quantified predicates, however, limitations of current SMT solvers in nonlinear arithmetic hinders them from being practically used in our framework, as demonstrated later in Sect.~\ref{sec_experiments}.
%where we will also pose several potential solutions that are expected to significantly reduce computational efforts induced by QE.

\subsubsection{Counterexample generation.}
If a candidate interpolant cannot be verified as an actual one, then at least one
witness can be found as a counterexample to that candidate, which
can be added to the set of sample points in the next iteration to refine further candidates (cf. Fig.~\ref{fig_nil_core}). Multiple counterexamples can be obtained at a time thereby effectively reducing the number of future iterations. 

In general, we have little control over which counterexample will be
returned by QE. In the worst case, the
counterexample can lie almost exactly on the hyperplane found by
SVM. This poses issues for the termination of the algorithm. We will
address this theoretical issue by slightly modifying the algorithm, as
explained in Sect.~\ref{subsec_variants} and~\ref{sec_theories}.

\subsection{Comparison with the Na\"ive QE-Based Method}

Simply performing QE on $\exists \yy.\ \phi(\xx, \yy)$ yields already an interpolant for mutually contradictory $\phi$ and $\psi$. Such an interpolant is actually the \emph{strongest} in the sense of~\cite{SPWK10}, which presents an ordered family of interpolation systems due to the logical strength of the synthesized interpolants. Dually, the negation of the result when performing QE over $\exists \zz.\ \psi(\xx, \zz)$ is the \emph{weakest} interpolant. However, as argued by D'Silva et al. in~\cite{SPWK10}, a good interpolant (approximation of $\phi$ or $\psi$) --when computing invariants of transition systems using interpolation-based model checking-- should be coarse enough to enable rapid convergence but strong enough to be contained within the weakest inductive invariant. In contrast, the advantages of NIL are two-fold: first, it produces better interpolants (in the above sense) featuring ``medium'' strength (due to the way optimal-margin classifier is defined) which are thus more effective in practical use and furthermore resilient to perturbations in $\phi$ and $\psi$ (i.e., the robustness shown later in Sect.~\ref{subsec_robustness}); second, NIL always returns a single polynomial inequality as the interpolant which is often simpler than that derived from the na\"ive QE-based method, where the direct projection of $\phi(\xx, \yy)$ onto the common space over $\xx$ can be as complex as the original $\phi$.

These issues can be avoided by combining this method with a template-based approach, which in turn introduces fresh quantifiers over unknown parameters to be eliminated. Note that in NIL the candidate interpolants $I\in \RR[\xx]_m$ under verification are polynomials without unknown parameters, and therefore, in contrast to performing QE over an assumed template, the learning-based technique can practically generate polynomial interpolants of higher degrees (with acceptable rounds of iterations). For example, NIL is able to synthesize an interpolant of degree 7 over 2 variables (depicted later in Fig.~\subref*{fig:ultimate}), which would require a polynomial template with $\tbinom{7+2}{2} = 36$ unknown parameters that goes far beyond the capability of QE procedures.

On the other hand, performing QE within every iteration of the learning process, for entailment checking and generating counterexamples, limits the efficiency of the proposed method, thereby confining NIL currently to applications only of small scales. Potential solutions to the efficiency bottleneck will be discussed in Sect.~\ref{sec_experiments}.
% that are expected to reduce computational efforts induced by QE.

\subsection{Variants of NIL}\label{subsec_variants}

While the above basic algorithm is already effective in practice (as demonstrated in Sect.~\ref{sec_experiments}), it
is guaranteed to terminate only when there is an interpolant with
positive functional margin between $\llbracket \phi \rrbracket$ and $\llbracket \psi \rrbracket$. In this section, we present two variants of the algorithm that have nicer
theoretical properties in cases where the two sets are only
separated by an interpolant with zero functional margin, e.g., cases where $\llbracket \phi \rrbracket$ and $\llbracket \psi \rrbracket$ share adjacent or even coincident boundaries.

\subsubsection{Entailment checking with tolerance $\delta$.}

When performing entailment checking for a candidate interpolant $I$,
instead of using, e.g., the formula $p(\xx)\ge 0$ for $I$, we can introduce a
tolerance of $\delta$. That is, we check the satisfiability of $\phi
\wedge (p(\xx)<-\delta)$ and $(p(\xx)\ge\delta) \wedge \psi$ instead
of the original $\phi \wedge (p(\xx)<0)$ and $(p(\xx)\ge
0)\wedge\psi$. This means that a candidate that is an interpolant ``up
to a tolerance of $\delta$'' will be returned as a true interpolant,
which may be acceptable in some applications. If the candidate
interpolant is still not verified, the counterexample is guaranteed to
be at least a distance of $\delta$ away from the separating
hyperplane. Note the distance $\delta$ is taken in the feature space
$\mathbb{R}^{\tilde{d}}$, not in the original space. We let
NIL$_\delta$($\phi$,$\psi$,$m$) denote the version of NIL with this
modification (cf. Fig.~\ref{fig_nil_delta}). In the next section, we show
NIL$_\delta$($\phi$,$\psi$,$m$) terminates as long as $\llbracket \phi
\rrbracket$ and $\llbracket \psi \rrbracket$ are bounded, including
the case where they are separated only by interpolants of functional
margin zero.

\subsubsection{Varying tolerance during the execution.}

A further refinement of the algorithm can be made by varying the
tolerance $\delta$ during the execution. We also introduce a bounding
box $B$ of the varying size to handle unbounded cases. Define algorithm
NIL$^*_{\delta,B}(\phi,\psi,m)$ as follows. Let $\delta_1=\delta$ and
$B_1=B$. For each iteration $i$, execute the core algorithm,
except that the counterexample must be a distance of at least
$\delta_i$ away from the separating boundary, and have absolute value
in each dimension at most $B$ (both in $\mathbb{R}^{\tilde{d}}$). After
the termination of iteration $i$, begin iteration $i+1$ with
$\delta_{i+1}=\delta_i/2$ and $B_{i+1}=2B_i$. This continues until an
interpolant is found or until a pre-specified cutoff. For any
$\llbracket \phi \rrbracket$ and $\llbracket \psi \rrbracket$ (without
the boundedness condition), this variant of the algorithm
\emph{converges} to an interpolant in the limit, which will be made
precise in the next section.
%
%\paragraph{Rational recovery.\chenSide{postponed to the implementation section.}}

\section{Soundness, Completeness and Convergence}\label{sec_theories}

In this section, we present theoretical results obtained for the basic NIL algorithm and its variants. Proofs are included in Appx.~\ref{appendix_proofs} due to the lack of space.

First, the basic algorithm is sound, as captured by
Theorem~\ref{theorem_sound}.

\begin{theorem}[Soundness of NIL]
	If NIL($\phi$,$\psi$,$m$) terminates and returns $I$, then $I$ is an interpolant in $\RR[\xx]_m$ of $\phi$ and $\psi$.
	\label{theorem_sound}
\end{theorem}

Under certain conditions, the algorithm is also terminating (and hence
complete). We prove two such situations below. In both cases, we
require boundedness of the two sets that we want to separate. In the
first case, there exists an interpolant with positive functional
margin between the two sets.

\begin{theorem}[Conditional Completeness of NIL]
  If $\llbracket \phi \rrbracket$ and $\llbracket \psi \rrbracket$ are
  bounded and there exists an interpolant in $\RR[\xx]_m$ of $\phi$ and $\psi$ with positive functional margin $\gamma$ when mapped to $\mathbb{R}^{\tilde{d}}$, then NIL($\phi$,$\psi$,$m$)
  terminates and returns an interpolant $I$ of $\phi$ and $\psi$.
\label{theorem_complete}
\end{theorem}

The standard algorithm is not guaranteed to terminate when $\llbracket
\phi \rrbracket$ and $\llbracket \psi \rrbracket$ are only separated
by interpolants of functional margin zero. However, the modified
algorithm NIL$_\delta$($\phi$,$\psi$,$m$) does terminate (with the
cost that the resulting answer is an interpolant with tolerance
$\delta$).

\begin{theorem}[Completeness of NIL$_\delta$ with zero margin]
  If $\llbracket \phi \rrbracket$ and $\llbracket \psi \rrbracket$ are
  bounded, and $\delta>0$, then NIL$_\delta$($\phi$,$\psi$,$m$)
  terminates. It returns an interpolant $I$ of $\phi$ and $\psi$ with
  tolerance $\delta$ whenever such an interpolant exists.
\label{theorem_complete2}
\end{theorem}

By iteratively decreasing $\delta$ during the execution of the
algorithm, as well as introducing an iteratively increasing bounding
box, as in NIL$^*_{\delta,B}(\phi,\psi,m)$, we can obtain more and
more accurate candidate interpolants. We now show that this algorithm
\emph{converges} to an interpolant without restrictions on $\phi$ and
$\psi$. We first make this convergence property precise in the
following definition.

\begin{definition}[Convergence of a sequence of equations to an interpolant]
  \label{def_convergence}
  Given two sets $\llbracket \phi \rrbracket$ and $\llbracket \psi
  \rrbracket$ that we want to separate, and an
  infinite sequence of equations $I_1,I_2,\dots$, we say the sequence $I_n$
  converges to an interpolant of $\phi$ and $\psi$ if, for each point
  $p$ in the interior of $\llbracket \phi
  \rrbracket$ or $\llbracket \psi
  \rrbracket$, there exists some integer
  $K_p$ such that $I_k$ classifies $p$ correctly for all $k\ge K_p$.
\end{definition}

\begin{theorem}[Convergence of NIL$^*_{\delta,B}$]
  Given two regions $\llbracket \phi \rrbracket$ and $\llbracket \psi
  \rrbracket$. Suppose there exists an interpolant of $\phi$ and
  $\psi$, then the infinite sequence of candidates produced by
  NIL$^*_{\delta,B}(\phi,\psi,m)$ converges to an interpolant in the
  sense of Definition \ref{def_convergence}.
  \label{theorem_convergence}
\end{theorem}

%\begin{proof}
%	Sketch: there exists an $m$-polynomial interpolant of $\phi$ and $\psi$. \\
%	$\Longrightarrow$ $X^+$ and $X^-$ are linearly separable in the $\left(\tbinom{n}{m+n}-1\right)$-dimensional feature space, where $n$ is the number of common variables of $\phi$ and $\psi$.\\
%	$\Longrightarrow$ by Lemma \ref{lemma_SVMs_corr}, SVMs produce a half-space $h$ s.t. $\forall \vec{x} \in X^+.\ h(\vec{x})>0$ and $\forall \vec{x} \in X^-.\ h(\vec{x})<0$.\\
%	The proof depends further on the convergence of the separating hyperplane $h(\xx)>0$ to a real interpolant $I$, as $X^+ \rightarrow \llbracket \phi \rrbracket$ and $X^- \rightarrow \llbracket \psi \rrbracket$. \textcolor{blue}{This however seems hard to prove...}
%	\qed
%\end{proof}

\section{Implementation and Experiments}\label{sec_experiments}

\subsection{Implementation Issues}

We have implemented the core algorithm NIL as a prototype\footnote{Available at \url{http://lcs.ios.ac.cn/~chenms/tools/NIL.tar.bz2}} in Wolfram Mathematica with LIBSVM \cite{DBLP:journals/tist/ChangL11} being integrated as an engine to perform SVM classifications. Despite featuring no completeness for adjacent $\llbracket \phi \rrbracket$ and $\llbracket \psi \rrbracket$ nor convergence for unbounded $\llbracket \phi \rrbracket$ or $\llbracket \psi \rrbracket$, the standard NIL algorithm yields already promising results as shown later in the experiments. Key Mathematica functions that are utilized include \textsc{Reduce}, for entailment checking, e.g., the unsatisfiability checking of $\phi \land \psi$ and the post-verification of a candidate interpolant, and \textsc{FindInstance}, for generating counterexamples and sampling initial points (when the random sampling strategy fails). The \textsc{Reduce} command implements a decision procedure for $\PT$ and its appropriate extension to catering for transcendental functions (cf.~\cite{Strzebonski11}) based on \emph{cylindrical algebraic decomposition} (CAD), due to Collins~\cite{Collins1975}. The underlying quantifier-elimination procedure, albeit inducing rather high computation complexity, cannot in practice be replaced by SMT-solving techniques (by checking the negation of a universally quantified predicate) as in the linear arithmetic. For instance, the off-the-shelf SMT solver Z3 fails to accomplish our tasks particularly when the coefficients occurring in the entailment problem to be checked get larger\footnote{As can be also observed at \url{https://github.com/Z3Prover/z3/issues/1765}}.

\subsubsection{Numerical errors and rounding.}

LIBSVM conducts floating-point computations for solving the optimization problems induced by SVMs and consequently yields numerical errors occurring in the candidate interpolants. Such numerical errors may block an otherwise valid interpolant from being verified as an actual one and additionally bring down the simplicity and thereby the effectiveness of the synthesized interpolant, thus not very often proving humans with clear-cut understanding. This is a common issue for approaches that reduce the interpolation problem to numerical solving techniques, e.g. SDP solvers exploited in~\cite{DXZ13,GDX16,DBLP:conf/aplas/OkudonoNKSKH17}, while an established method to tackle it is known as \emph{rational recovery}~\cite{lang2012introduction,Zhang2007}, which retrieves the nearest rational number from the continued fraction representation of its floating-point approximation at any given accuracy (see e.g.~\cite{Zhang2007} for theoretical guarantees and~\cite{DBLP:conf/aplas/OkudonoNKSKH17} for applications in interpolation). The algorithm implementing rational recovery has been integrated in our implementation and the consequent benefits are two-fold: (i) NIL can now cope with interpolation tasks where only exact coefficients suffice to constitute an actual interpolant while any numerical error therein will render the interpolant invalid, e.g., cases where $\llbracket \phi \rrbracket$ and $\llbracket \psi \rrbracket$ share parallel, adjacent, or even coincident boundaries, as demonstrated later by examples with ID 10--17 in Table~\ref{tb:benchmark}; (ii) rationalizing coefficients moreover facilitates simplifications over all of the candidate interpolants and therefore practically accelerating the entailment checking and counterexample generation processes, which in return yields simpler interpolants, as shown in Table~\ref{tb:comparison} in the following section.
%
%\SetAlgoRefName{Rational Recovery}
%\begin{algorithm}[h]
%	\caption{Rational Recovery of Interpolants}\label{alg_recovery}
%	\SetKwInOut{Input}{input}\SetKwInOut{Output}{output}\SetNoFillComment
%	\Input{$\phi$ and $\psi$ in $\PT$; \ $I$, interpolant learned by SVM; $p_{max}$, maximum precision.
%	}
%	\tcc{rational recovery of the interpolant}
%	\tcc{precision initialization}
%	$p \leftarrow 1$\;
%	\While{$p \ge p_{max}$\label{alg_recovery_loop}}
%	{
%		\tcc{round the coefficients of polynomial interpolants to precision $p$}
%		$I' \leftarrow \textbf{Round}(I)$\; 
%		\tcc{checking the validity of rounded interpolant}
%		\If{$\phi \models I \ \mathrm{and}\ I \wedge \psi \models \bot$\label{alg_recovery_checking}}
%		{
%			\Return $I$\;\label{alg_recovery_return}
%		}
%		\tcc{elevate the precision}
%		\Else
%		{
%			$p \leftarrow p / 10$\;\label{alg_recovery_acc}
%		}
%	}
%	\tcc{abort if we cannot find the rational recovered interpolant}
%	\textbf{abort}\;
%\end{algorithm} 
%
%\chen{why not SMT(Z3); why Mathematica (Reduce/FindInstance); sampling; rational recovery (rounding).}

\subsection{Benchmark and Experimental Results}

Table~\ref{tb:benchmark} collects a group of benchmark examples from the literature on synthesizing nonlinear interpolants as well as some geometrically contrived ones. All of the experiments have been evaluated on a 3.6GHz Intel Core-i7 processor with 8GB RAM running 64-bit Ubuntu 16.04.

\begin{figure}[!t]
	\centering
	\hspace*{-.52cm}
	\begin{tabular}{ccc}
		\subfloat[Adjacent]{~~~~\includegraphics[scale=0.28]{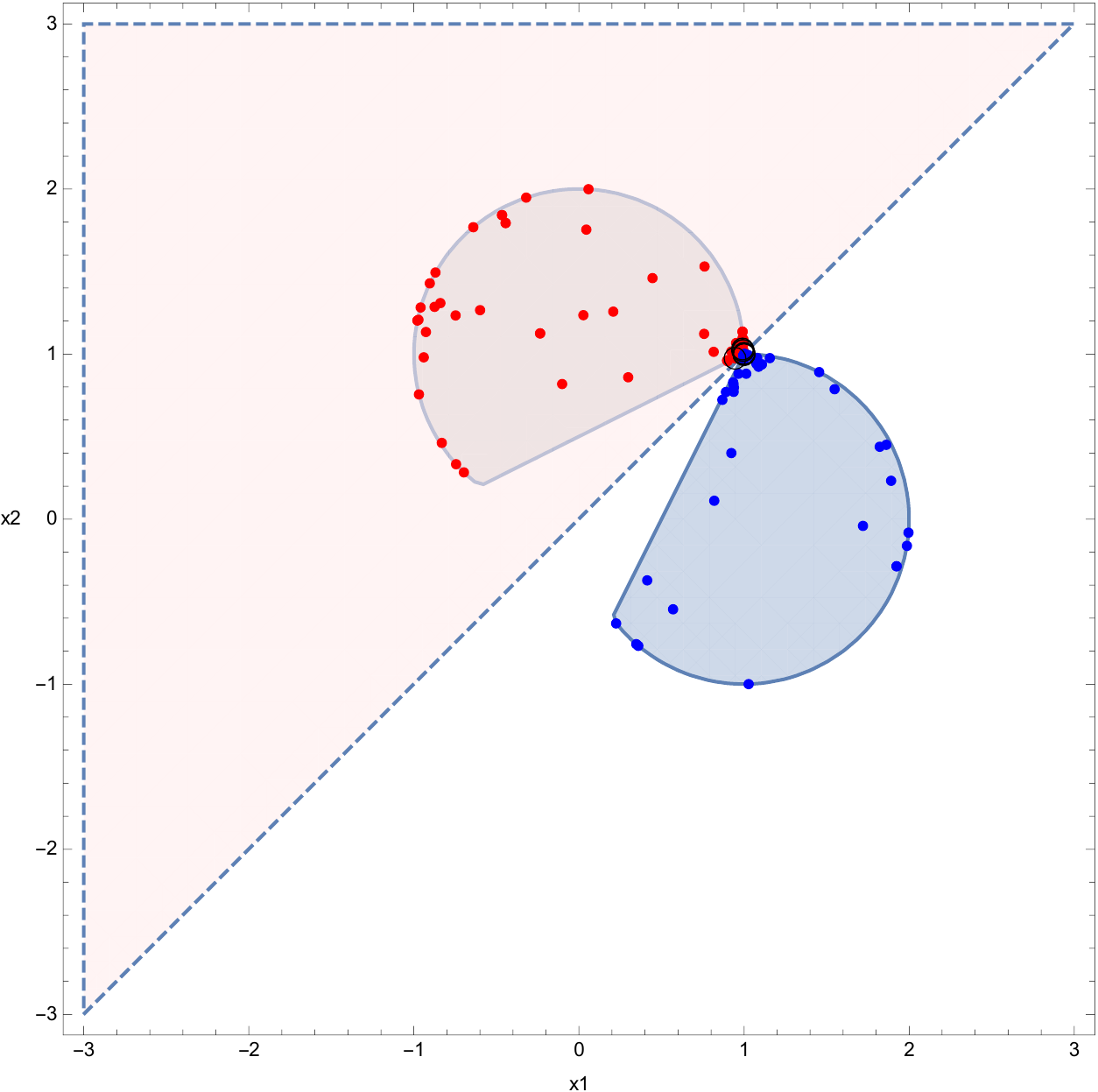}~~~~\label{fig:IJCAR16}}& %$\quad$
		\subfloat[Ultimate]{~~~~\includegraphics[scale=0.28]{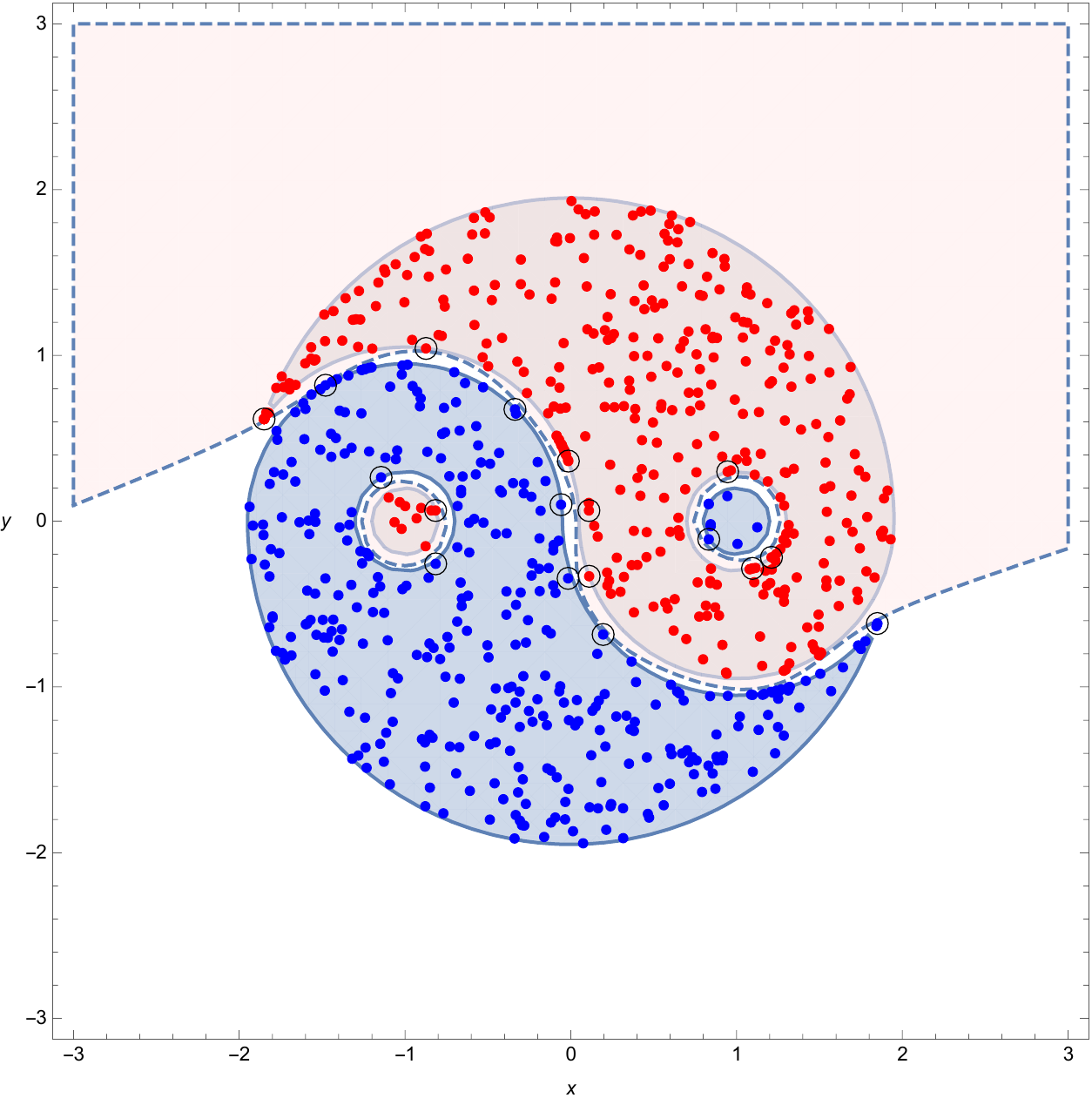}~~~~\label{fig:ultimate}}& %$\quad$
		\subfloat[Twisted]{~~~~\includegraphics[scale=0.28]{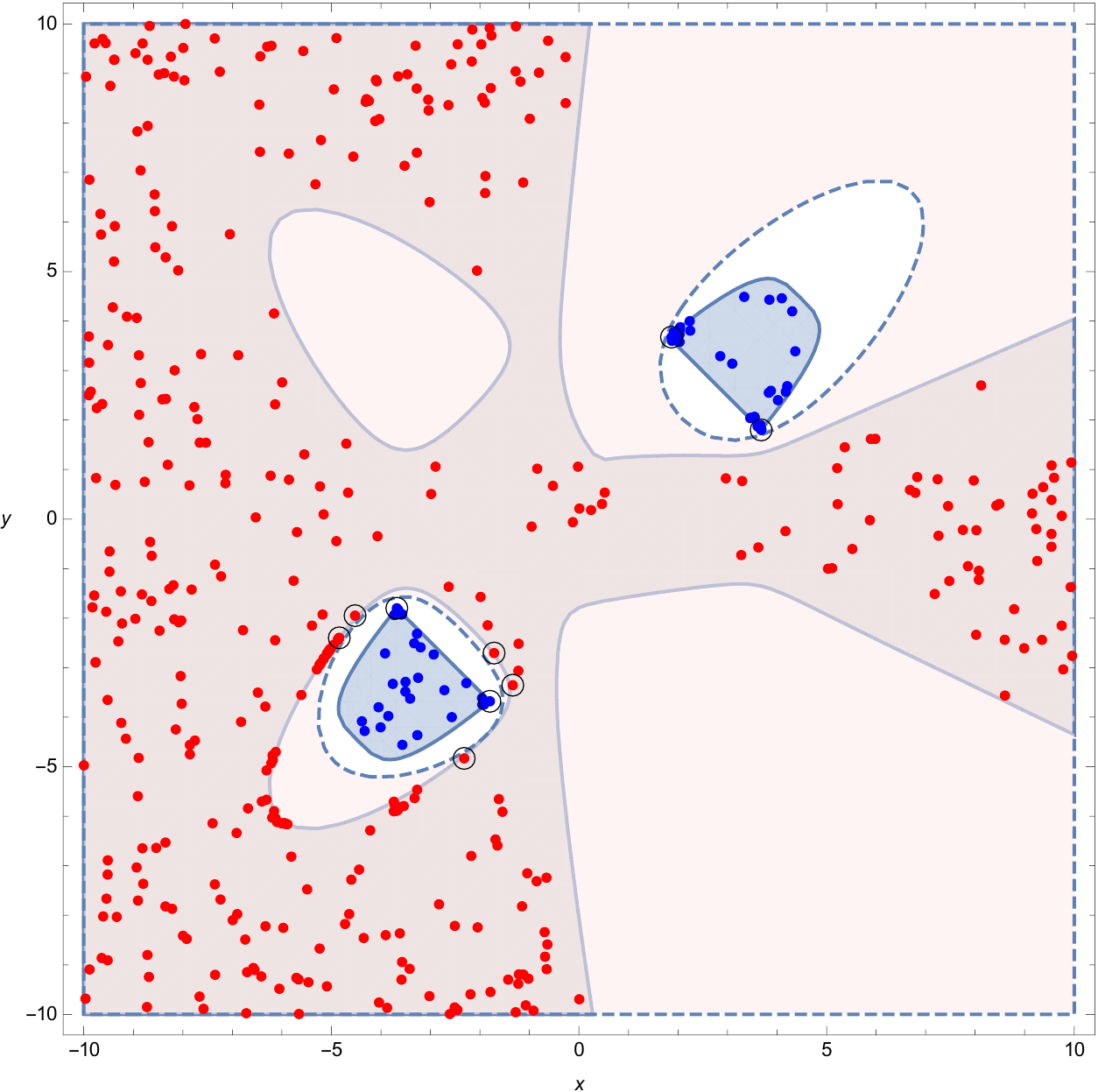}~~~~\label{fig:twisted}}\\[-.1cm]
		\subfloat[Parallel parabola]{~~~~\includegraphics[scale=0.28]{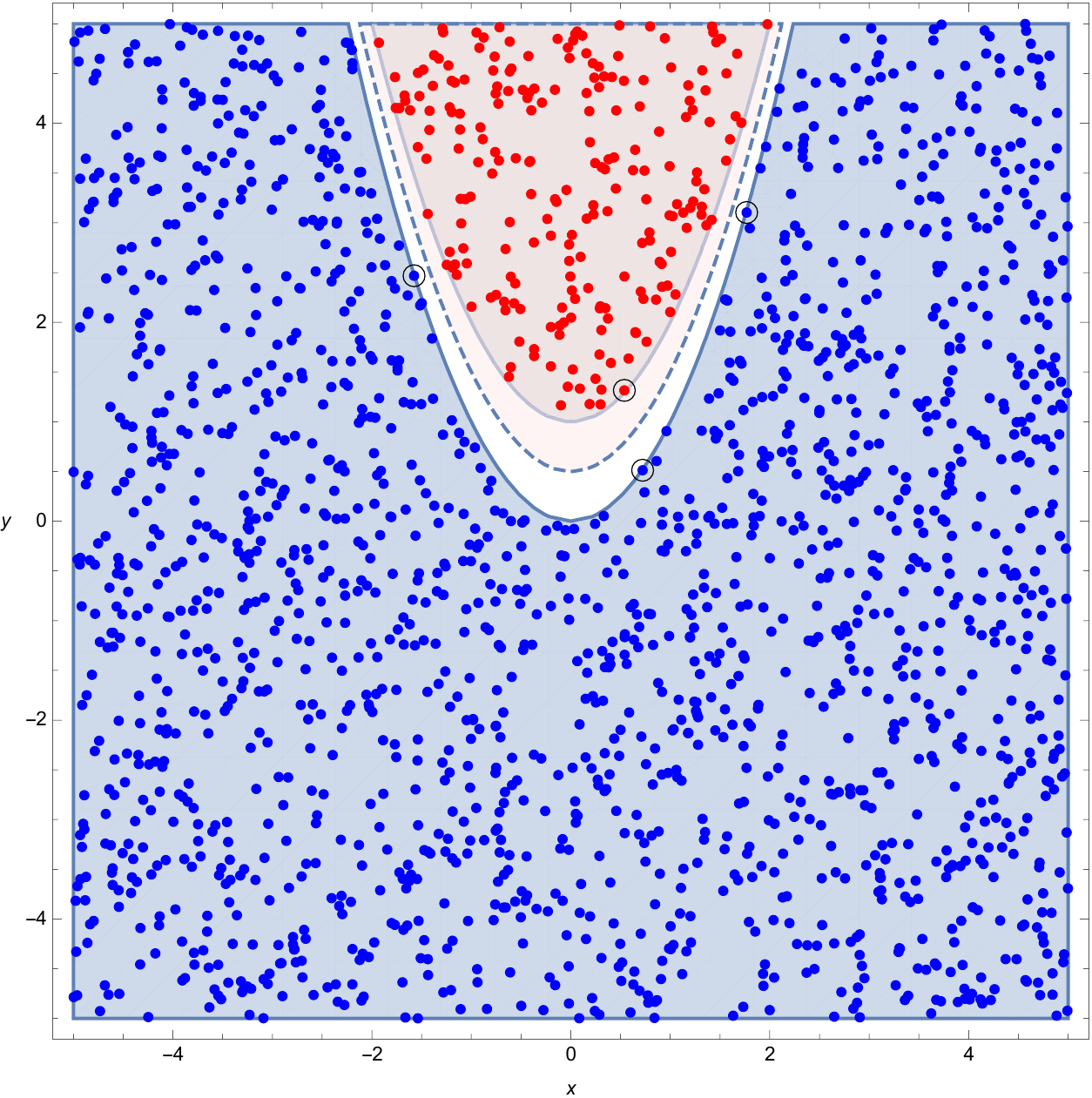}~~~~\label{fig:parallel}}& 
		%$\quad$
		\subfloat[Coincident]{~~~~\includegraphics[scale=0.28]{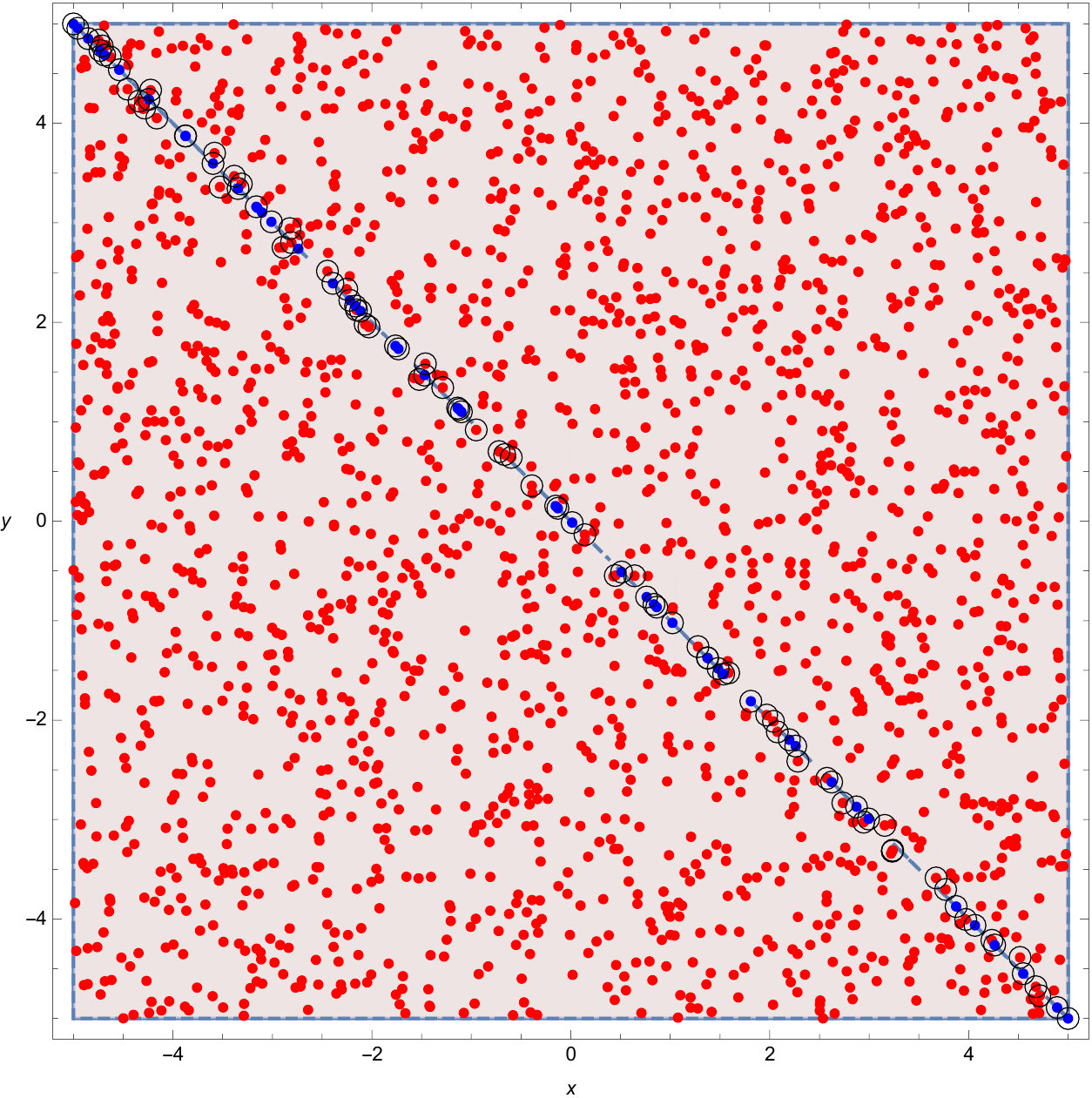}~~~~\label{fig:adjacent}}& %$\quad$
		\subfloat[Sharper-2]{~~~~\includegraphics[scale=0.28]{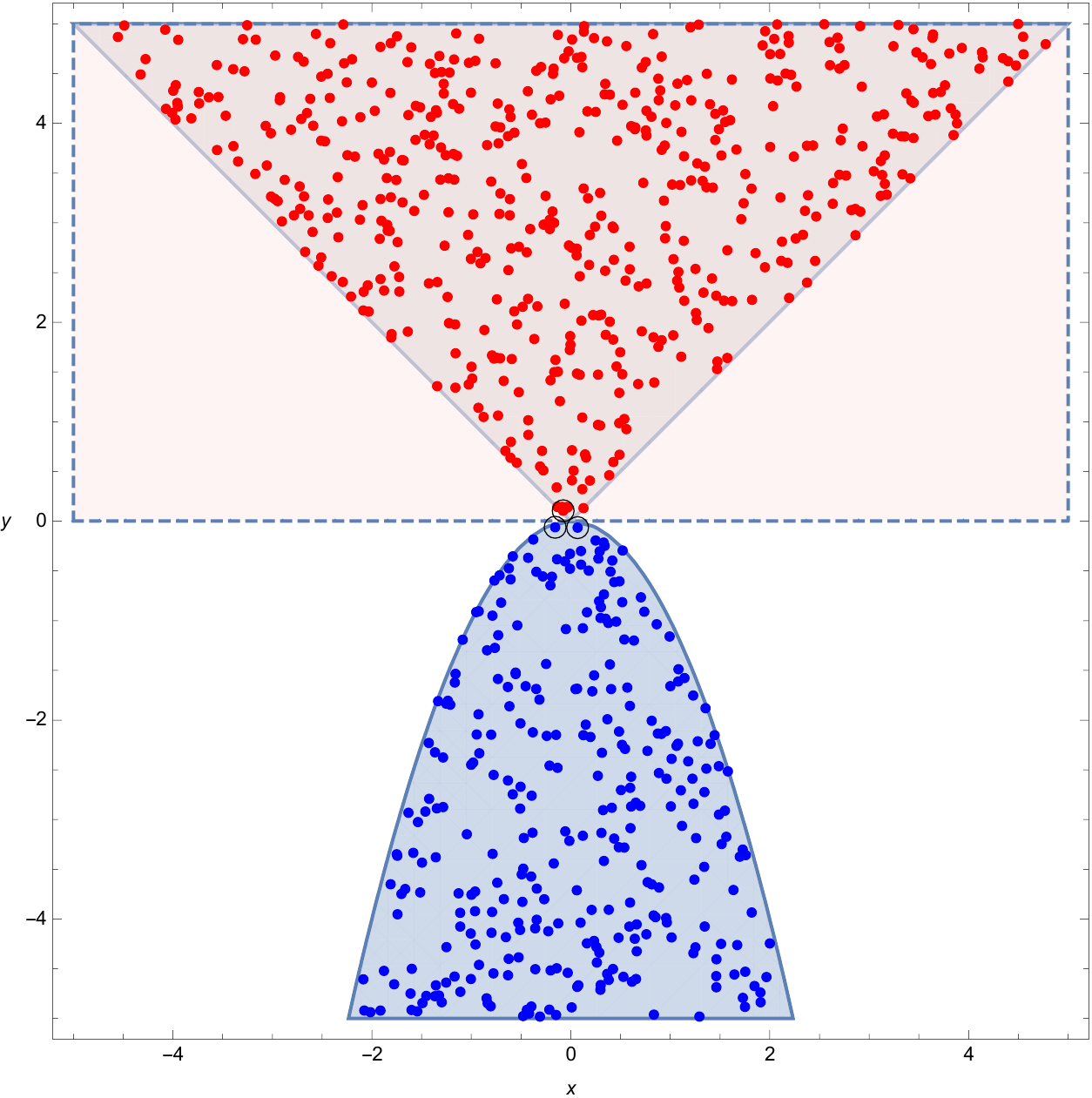}~~~~\label{fig:sharper}}\\[-.1cm]
		\subfloat[Face]{~~~~\includegraphics[scale=0.28]{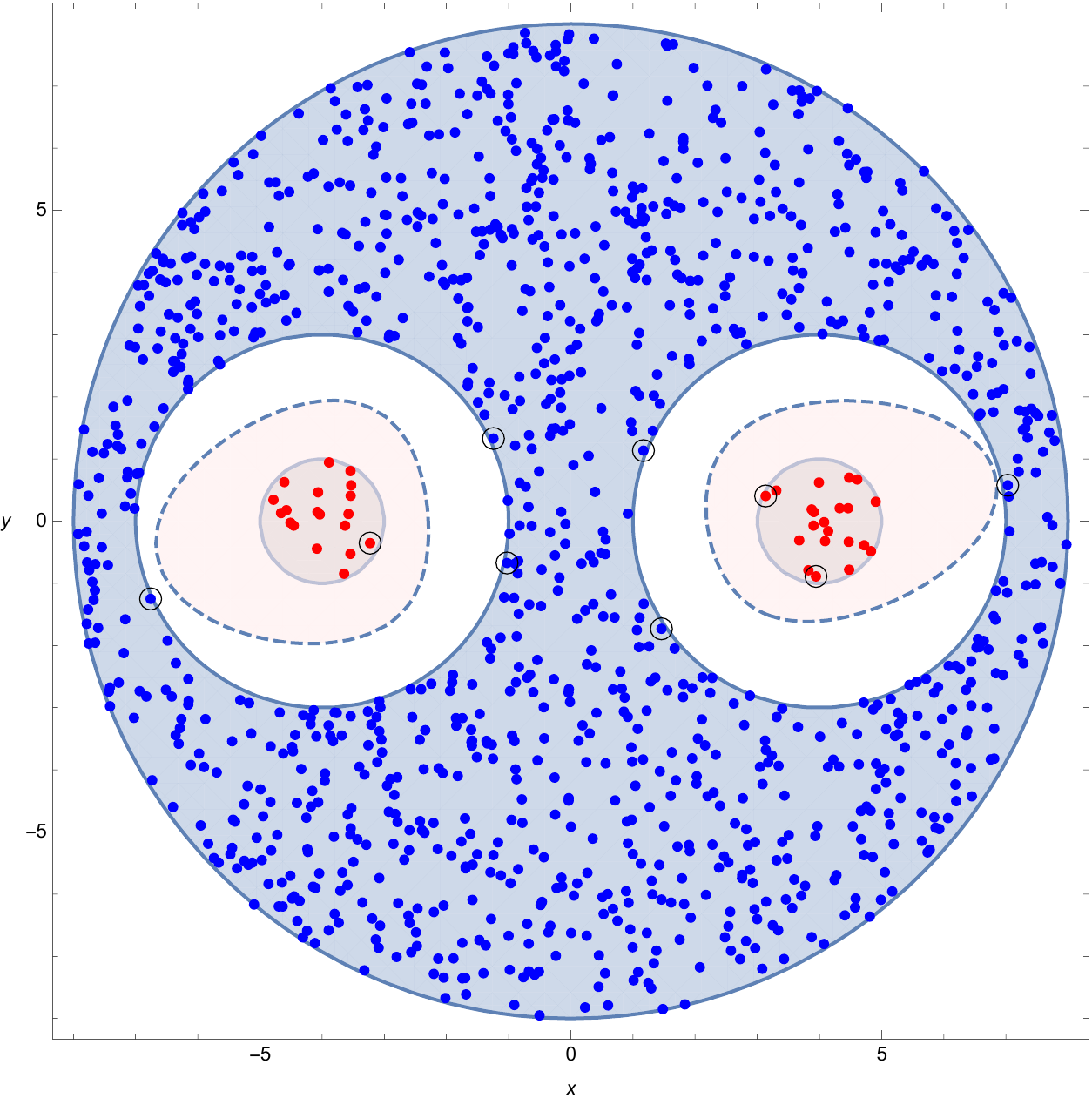}~~~~\label{fig:face}}& 
		%$\quad$
		%\subfloat[$\epsilon$-perturbations in the radii]{~~~~\scalebox{0.313}{\themybox}~~~~\label{fig:face-epsilon-sketch}} &
		%%$\quad$
		%\subfloat[Face-$\epsilon$]{~~~~\includegraphics[scale=0.3]{face-epsilon}~~~~\label{fig:face-epsilon}}\\
		\hspace*{.1cm}
		\subfloat[CAV13-2]{~~~~\includegraphics[scale=0.28]{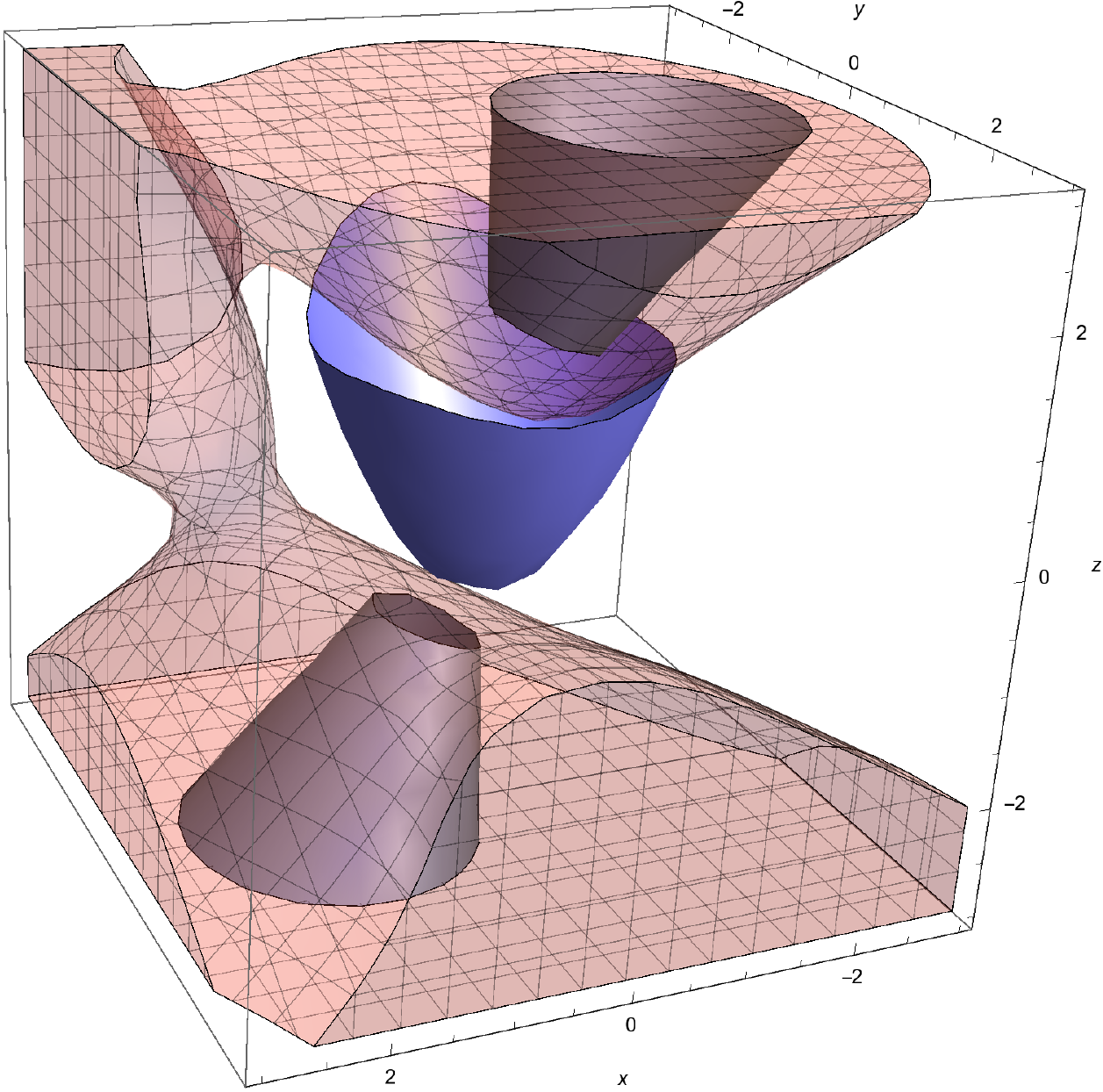}~~~~\label{fig:CAV13}}
		\hspace*{-.1cm}& 
		%$\quad$
		\subfloat[TACAS16]{~~~~\includegraphics[scale=0.28]{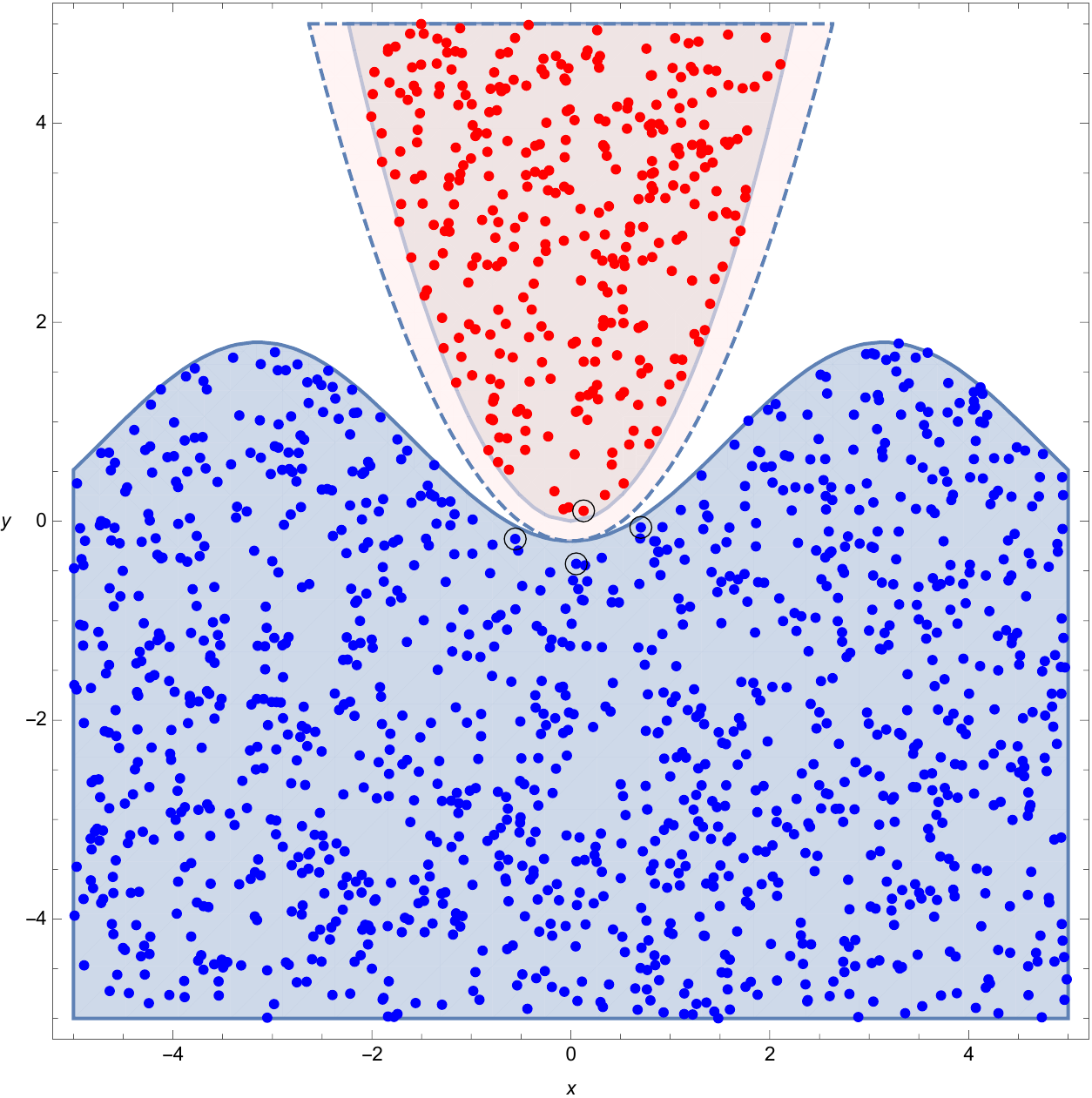}~~~~\label{fig:transcendental}}
	\end{tabular}
	\caption{Visualization in NIL on a selected set of examples. Legends: gray region: $\llbracket \phi \rrbracket$, blue region: $\llbracket \psi \rrbracket$, pink region: $\llbracket I \rrbracket$ with a valid interpolant $I$, red dots: $X^+$, blue dots: $X^-$, circled dots: support vectors. Sample points are hidden in 3D-graphics for a clear presentation.}\label{fig:visualization}
\end{figure}

In Table~\ref{tb:benchmark}, we group the set of examples into four categories comprising 20 cases in total. For each example, \textbf{ID} numbers the case, $\bm{\phi}$, $\bm{\psi}$ and $\bm{I}$ represent the two formulas to be interpolated and the synthesized interpolant by our method respectively, while \textbf{Time/s} indicates the total time in seconds for interpolation. The categories are described as follows, and the visualization of a selected set of typical examples thereof is further depicted in Fig.~\ref{fig:visualization}.
\begin{description}[align=left, leftmargin=0pt, itemsep=2.2pt]
\item[Cat. I: with/without rounding.] This category includes 9 cases, for which our method generates the polynomial interpolants correctly with or without the rounding operation.
\item[Cat. II: with rounding.] For cases 10 to 17 in this category, where $\llbracket \phi \rrbracket$ and $\llbracket \psi \rrbracket$ share parallel, adjacent, or even coincident boundaries, our method produces interpolants successfully with the rouding process based on rational recovery.
\item[Cat. III: beyond polynomials.] This category encloses two cases beyond the theory $\PT$ of polynomials: for case 18, a verified polynomial interpolant is obtained in spite of the transcendental term in $\psi$; while for case 19, the SVM classification fails since $\llbracket \phi \rrbracket$ and $\llbracket \psi \rrbracket$ are not linearly separable in any finite-dimensional feature space and hence no polynomial interpolant exists for this example. Note that our counterexample-guided learning framework admits a straightforward extension to a decidable fragment of more general nonlinear theories involving transcendental functions, as investigated in~\cite{Strzebonski11}.
\item[Cat. IV: unbalanced.] The case 20, called Unbalanced, instantiates a particular scenario where $\phi$ and $\psi$ have extraordinary ``unbalanced'' number of models that make them true respectively. For this example, there are an infinite number of models satisfying $\phi$ yet one single model (i.e., $x=0$) satisfying $\psi$. The training process in SVMs may fail when encountering extremely unbalanced number of positive/negative samples. This is solved by specifying a \emph{weight} factor for the positive set of samples as the number of negative ones, and dually for the other way around, to balance biased number of training samples before triggering the classification. Such a balancing trick is supported in LIBSVM.
\end{description}%
Remark that examples named CAV13-1/3/4 are taken from~\cite{DXZ13} (and the latter two originally from~\cite{KupferschmidB11} and~\cite{DBLP:conf/tacas/GulavaniCNR08} respectively), where interpolation is applied to discovering \emph{inductive invariants} in the verification of programs and hybrid systems. For instance, CAV13-3 is a program fragment describing an accelerating car and the synthesized interpolant by NIL suffices to prove the safety property of the car concerning its velocity.
%We refer the readers to~\cite{KupferschmidB11} and references therein for detailed explanations.

%\begin{landscape}
\begin{sidewaystable}[ph!]
%\begin{table}
	\caption{Benckmark examples for synthesizing nonlinear interpolants.}
	\label{tb:benchmark}
	\vspace*{-.2cm}
	\centering
	\tiny
	\resizebox{.85\textwidth}{!}{
	\begin{tabular}{lllllll}%{llll|l|l|l}
		\toprule
		\textbf{Category} & \textbf{ID} & \textbf{Name} & $\bm{\phi}$ & $\bm{\psi}$ & $\bm{I}$ & \textbf{Time/s} \\
		\midrule
		\multirow{9}*{\vspace*{-9.9cm}\vtop{\hbox{\strut\bf with/without}\hbox{\strut\bf rounding}}}
		& 1 & Dummy & $x < -1$ & $x \ge 1$ & $x < 0$ & 0.11 \\
		& 2 & Necklace & $y-x^2-1=0$ & $y+x^2+1=0$ & $-y<0$ & 0.21 \\
		& 3 & Face & $\begin{aligned} &(x+4)^2 + y^2 - 1 \le 0 \vee \\ &(x-4)^2 + y^2 - 1 \le 0 \end{aligned}$ & $\begin{aligned} &x^2 + y^2 - 64 \le 0 \wedge \\ &(x+4)^2 + y^2 - 9 \ge 0 \wedge \\ &(x-4)^2 + y^2 - 9 \ge 0 \end{aligned}$ & $\begin{aligned}&\frac{x^4}{223}-\frac{x^3y}{356}+x^2(\frac{y^2}{45}-\frac{y}{170}-\frac{2}{9})+\\&x(\frac{y^3}{89}+\frac{y^2}{68}-\frac{y}{74}-\frac{1}{55})+\frac{y^4}{146}+\\&\frac{y^3}{95}+\frac{y^2}{37}+\frac{y}{366}+1<0\end{aligned}$ & 0.33 \\
		& 4 & Twisted & $\begin{aligned}&x^2-2 x y^2+3 x z-y^2 \\& -y z+z^2-1\geq 0\land \\ &\frac{1}{120} \left(-x^6-y^6\right)+x^2 z^2-  \\& x^2+\frac{1}{6} \left(x^4+2 x^2 y^2+y^4\right)+ \\&y^2 z^2-y^2-4\leq 0 \end{aligned}$ & $\begin{aligned}&w^2+4(x-y)^4+(x+y)^2-80\leq 0\land \\ &-w^2 (x-y)^4+100 (x+y)^2-3000\geq 0 \end{aligned}$ & $\begin{aligned}&-\frac{x^4}{160}+x^3 \left(\frac{y}{170}-\frac{1}{113}\right)+x^2\left(-\frac{y^2}{225}+\frac{y}{76}+\frac{2}{27}\right)+\\&x\left(\frac{y^3}{259}+\frac{y^2}{63}+\frac{5y}{51}-\frac{1}{316}\right)-\frac{y^4}{183}-\frac{y^3}{94}+\frac{y^2}{14}+\frac{y}{255}-1<0
		\end{aligned}$ & 140.62 \\
		& 5 & Ultimate & $\begin{aligned} &(x^2+y^2-3.8025\leq 0\land y\geq 0 \lor \\&(x-1)^2+y^2-0.9025\leq0) \land \\& (x-1)^2+y^2-0.09>0 \land \\&(x+1)^2+y^2-1.1025\geq 0 \lor \\&(x+1)^2+y^2-\frac{1}{25}\leq 0\end{aligned}$ & $\begin{aligned} &(-3.8025 + x^2 + y^2 \leq 0 \land -y \geq 0 \lor \\&-0.9025 + (-1 - x)^2 + y^2 \leq 0) \land \\& -0.09 + (-1 - x)^2 + y^2 > 0 \land \\&-1.1025 + (1 - x)^2 + y^2 \geq 0 \lor \\&-\frac{1}{25} + (1 - x)^2 + y^2 \leq 0 \end{aligned}$ & $\begin{aligned}&\frac{x^7}{27}+x^6(-\frac{y}{5}-\frac{1}{96})+x^5(\frac{2y^2}{9}-\frac{y}{32}-\frac{1}{2})+\\&x^4(-\frac{2y^3}{9}+\frac{y}{3}+\frac{1}{31})+x^3(\frac{y^4}{11}+\frac{y^3}{10}-\frac{10y^2}{13}+\frac{y}{18}+\frac{15}{16})+\\&x^2(-\frac{y^5}{25}-\frac{y^4}{18}-\frac{y^3}{3}+\frac{y^2}{10}-\frac{1}{32})+\\&x\left(\frac{y^6}{71}+\frac{2y^4}{11}-\frac{y^3}{25}-y^2-\frac{y}{45}-\frac{3}{8}\right)+\\&\frac{y^6}{48}-\frac{y^5}{7}+\frac{y^4}{6}-\frac{y^3}{2}-\frac{y^2}{6}-\frac{y}{59}+\frac{1}{85}<0\end{aligned}$ & 48.82 \\
		& 6 & IJCAR16-1~\cite{GDX16} & $\begin{aligned}&-x_1^2 + 4x_1 + x_2 - 4 \geq 0 \land \\& -x_1 - x_2 + 3 - y^2 > 0\end{aligned}$ & $-3x_1^2 - x_2^2 + 1 \geq 0 \land x_2 - z^2 \geq 0$ & $1-\frac{3x_1}{4}-\frac{x_2}{2}<0$ & 0.16 \\
		& 7 & CAV13-1~\cite{DXZ13} & $\begin{aligned}&1- a^2 - b^2 > 0 \land a^2 + b - 1 - x = 0 \land \\& b + bx + 1 - y = 0\end{aligned}$ & $x^2 - 2y^2 - 4 > 0$ & $-1+\frac{x^2}{2}-\frac{y}{3}+\frac{xy}{3}-\frac{y^2}{4}<0$ & 3.25 \\
		& 8 & CAV13-2~\cite{DXZ13} & $\begin{aligned}&x^2 + y^2 + z^2 - 2 \geq 0 \land \\& 1.2 x^2 + y^2 + x z = 0\end{aligned}$ & $\begin{aligned}&20 - 3x^2 - 4y^3 - 10z^2 \geq 0 \land \\&x^2 + y^2 - z - 1 = 0\end{aligned}$ & $\begin{aligned}&105 x^4+x^2 (140 y^2+24 y (5 z+7)+35 z (3 z+8))+ \\&
		2 (70 y^3 z+5 y^2 (12 z^2+21 z+28)-14 y (6 z^3+5 z^2+ \\&
		10)- 35 (3 z^4+8 z^2+4 z-9))<14 x (20 x^2 (z+1)+ \\&
		10 y^2 (z+2)- 3 y (4 z^2-5 z+4)-20 z (z^2+2))
		\end{aligned}$ & 3857.89 \\
		& 9 & CAV13-3~\cite{DXZ13} & $\begin{aligned}&vc < 49.61 \land fa = 0.5418 vc^2 \land \\& fr = 1000 - fa \land ac = 0.0005 fr \land \\& vc_1 = vc + ac\end{aligned}$ & $vc_1 \geq 49.61$ & $-1+\frac{2 vc_1}{99}<0$ & 40.63 \\
		\midrule
		\multirow{8}*{\vspace*{-1.5cm}\vtop{\hbox{\strut\bf with}\hbox{\strut\bf rounding}}}
		& 10 & Parallel parabola & $y - x^2 - 1 \geq 0$ & $y - x^2 < 0$ & $\frac{1}{2}+x^2<y$ & 4.50 \\
		& 11 & Parallel halfplane & $y - x - 1 \geq 0$ & $y - x + 1 < 0$ & $x<y$ & 2.46 \\
		& 12 & Sharper-1~\cite{DBLP:conf/aplas/OkudonoNKSKH17} & $y + 1 < 0$ & $x^2 + y^2 - 1 \leq 0$ & $2+y<y^2$ & 2.19  \\
		& 13 & Sharper-2~\cite{DBLP:conf/aplas/OkudonoNKSKH17} & $y - x > 0 \land x + y > 0$ & $y + x^2 <= 0$ & $y>0$ & 2.38 \\
		& 14 & Coincident & $x + y > 0 \lor x + y < 0$ & $x + y = 0$ & $(x+y)^2>0$ & 0.18 \\
		& 15 & Adjacent & $y - x^2 > 0$ & $y - x^2 <= 0$ & $x^2<y$ & 0.25 \\
		& 16 & IJCAR16-2~\cite{GDX16} & $\begin{aligned}&-y_1 + x_1 - 2 \geq 0 \land 2x_2 - x_1 - 1 > 0 \land \\&-y_1^2 - x_1^2 + 2x_1y_1 - 2y_1 + 2x_1 \geq 0 \land \\&-y_2^2 - y_1^2 - x_2^2 - 4y_1 + 2x_2 - 4 \geq 0\end{aligned}$ & $\begin{aligned}&-z_1 + 2x_2 + 1 \geq 0 \land 2x_1 - x_2 - 1 > 0\land \\&-z_1^2 - 4x_2^2 + 4x_2z_1 + 3z_1 - 6x_2 - 2 \geq 0 \land \\&-z_2^2 - x_1^2 - x_2^2 + 2x_1 + z_1 - 2x_2 - 1 \geq 0\end{aligned}$ & $x_1<x_2$ & 12.33 \\
		& 17 & CAV13-4~\cite{DXZ13} & $\begin{aligned}&xa_1 + 2 ya_1 \geq 0 \land xa_1 + 2 ya_1 - x_1 = 0 \land \\&-2 xa_1 + ya_1 - y_1 = 0 \land x - x_1 - 1 = 0 \land \\&y = y_1 + x \land xa = x - 2y \land ya = 2x + y\end{aligned}$ & $xa + 2 ya < 0$ & $2 xa + 4 ya>5$ & 3.10 \\
		\midrule
		\multirow{2}*{\vtop{\hbox{\strut\bf beyond}\hbox{\strut\bf polynomials}}}
		& 18 & TACAS16~\cite{DBLP:conf/tacas/GaoZ16} & $y - x^2 \geq 0$ & $y + \cos x - 0.8 \leq 0$ & $15 x^2<4+20 y$ & 12.71 \\
		& 19 & Transcendental & $\sin x \geq 0.6$ & $\sin x \leq 0.4$ & SVM failed & -- \\
		\midrule
		\multirow{1}*{\vtop{\hbox{\strut\bf unbalanced}}}
		& 20 & Unbalanced & $x > 0 \lor x < 0$ & $x=0$ & $x^2>0$ & 0.11 \\
		\bottomrule
	\end{tabular}
	}
%\end{table}
\end{sidewaystable}
\begin{table}[!h]
	\caption{Comparison of the synthesized interpolants.}
	\label{tb:comparison}
	\vspace*{.2cm}
	\hspace*{-.18cm}
	\centering
	\tiny
	\begin{tabular}{lllll}
		\toprule
		\textbf{Name}& &\textbf{Interpolants by NIL}& &\textbf{Interpolants from the sources}\\
		\midrule
		IJCAR16-1~\cite{GDX16}& & $1-\frac{3x_1}{4}-\frac{x_2}{2}<0$& &$-3+2x_1+x_1^2+\frac{1}{2}x_2^2>0$\\
		& &\\
		CAV13-1~\cite{DXZ13}& & $-1+\frac{x^2}{2}-\frac{y}{3}+\frac{xy}{3}-\frac{y^2}{4}<0$& &$436.45(x^2-2y^2-4) + \frac{1}{2}\le 0$\\
		& & & & \\
		%		CAV13-2~\cite{DXZ13}& &$\begin{aligned}&\frac{4}{3} \left(\frac{x^4}{8}+x^3 \left(-\frac{z}{3}-\frac{1}{3}\right)+x^2 \left(\frac{y^2}{6}+ y \left(\frac{z}{7}+\frac{1}{5}\right)+ \right. \right. \\& \left. \left. \frac{z^2}{8}+\frac{z}{3}\right)+ x \left(y^2 \left(-\frac{z}{6}-\frac{1}{3}\right)+y \left(\frac{z^2}{5}-\frac{z}{4}+\frac{1}{5}\right)+ \right. \right. \\& \left. \left. \frac{z^3}{3}+\frac{2 z}{3}\right)+\frac{y^3 z}{6}+y^2 \left(\frac{z^2}{7}+\frac{z}{4}+\frac{1}{3}\right)+y \left(-\frac{z^3}{5}- \right. \right. \\& \left. \left. \frac{z^2}{6}-\frac{1}{3}\right)- \frac{z^4}{4}-\frac{2 z^2}{3}-\frac{z}{3}+\frac{3}{4}\right)<0
		%		\end{aligned}$& &
		CAV13-2~\cite{DXZ13}& &$\begin{aligned}& 105 x^4+x^2 (140 y^2+24 y (5 z+7)+35 z (3 z+8))+ \\&
		2 (70 y^3 z+5 y^2 (12 z^2+21 z+28)-14 y (6 z^3+5 z^2+ \\&
		10)- 35 (3 z^4+8 z^2+4 z-9))<14 x (20 x^2 (z+1)+ \\&
		10 y^2 (z+2)- 3 y (4 z^2-5 z+4)-20 z (z^2+2))
		\end{aligned}$& & 
		$\begin{aligned}
		&-14629.26 + 2983.44x_3 + 10972.97x_3^2+\\& 297.62x_2 + 297.64x_2x_3 +	0.02x_2x^2_3 + 9625.61x_2^2-\\& 1161.80x_2^2x_3 +0.01x_2^2x^2_3 +811.93x^3_2 +\\& 2745.14x^4_2-10648.11x_1+3101.42x_1x_3+\\&8646.17x_1x^2_3+511.84x_1x_2-1034x_1x_2x_3 +\\& 0.02x_1x_2x_3^2+9233.66x_1x_2^2 + 1342.55x_1x_2^2x_3 -\\& 138.70x_1x_2^3 + 11476.61x^2_1 - 3737.70x_1^2x_3+\\&4071.65x_1^2x_3^2-2153.00x1_2x_2+373.14x_1^2x_2x_3+\\&7616.18x_1^2x_2^2 + 8950.77x_1^3+1937.92x_1^3x_3-\\&64.07x_1^3x_2+4827.25x_1^4 > 0
		\end{aligned}$\\
		& & & & \\
		CAV13-3~\cite{DXZ13}& &$-1+\frac{2 vc_1}{99}<0$& &$-1.3983vc_1+69.358>0$\\
		& & & & \\
		Sharper-1~\cite{DBLP:conf/aplas/OkudonoNKSKH17}& &$2+y<y^2$& &$34y^2-68y-102\ge 0$\\
		& & & & \\
		Sharper-2~\cite{DBLP:conf/aplas/OkudonoNKSKH17}& &$y>0$& &$8y+4x^2>0$\\
		& & & & \\
		IJCAR16-2~\cite{GDX16}& &$x_1<x_2$& &$-x_1+x_2>0$\\
		& & & & \\
		CAV13-4~\cite{DXZ13}& &$2 xa + 4 ya>5$& & $\begin{aligned}&716.77+1326.74(ya)+1.33(ya)^2+433.90(ya)^3+\\&668.16(xa)-155.86(xa)(ya)+317.29(xa)(ya)^2+\\&222.00(xa)^2+592.39(xa)^2(ya)+271.11(xa)^3>0\end{aligned}$\\
		& & & & \\
		TACAS16~\cite{DBLP:conf/tacas/GaoZ16}& &$15 x^2<4+20 y$& &$\begin{aligned}&y>1.8\lor(0.59\leq y\leq 1.8\land -1.35\leq x\leq 1.35)\lor\\& (0.09\leq y<0.59\land -0.77\leq x\leq0.77)\lor\\&(y\geq 0\land -0.3\leq x\leq 0.3)\end{aligned}$\\
		\bottomrule
	\end{tabular}
\end{table}

\subsubsection{Applicability and comparison with existing approaches.}

As shown in Table~\ref{tb:benchmark}, our learning-based technique succeeds in all of the benchmark examples that admit polynomial interpolants. Due to theoretical limitations of existing approaches as elaborated in Sect.~\ref{sec_intro}, none of the aforementioned methods can cope with as many cases in Table~\ref{tb:benchmark} as NIL can. For instances, the Twisted example as depicted in Fig.~\subref*{fig:twisted} falls beyond the scope of concave quadratic formulas and thus cannot be addressed by the approach in~\cite{GDX16}, while the Parallel parabola example as shown in Fig.~\subref*{fig:parallel} needs an infinite combination of linear constraints as an interpolant when performing the technique in~\cite{DBLP:conf/tacas/GaoZ16} and hence not of practical use, to name just a few. Moreover, we list in Table~\ref{tb:comparison} a comparison of the synthesized interpolants against works where the benchmark examples are collected from. As being immediately obvious from Table~\ref{tb:comparison}, our technique often produces interpolants of simpler forms, particularly for examples CAV13-2, CAV13-4 and TACAS16. Such a simplicity benefits from both the rounding effect and the form of interpolant (i.e., a single polynomial inequality) that we tend to construct.

\subsubsection{Bottleneck of efficiency and potential solutions.}

The current implementation of NIL works promisingly for small examples; it does not scale to interpolation problems with large numbers of common variables, as reported in Table~\ref{tb:benchmark}. The bottleneck stems from quantifier eliminations performed within every iteration of the learning process, for entailment checking and generating counterexamples. We pose here several potential solutions that are expected to significantly reduce computational efforts: (i) substitute general purpose QE procedure that perform CAD by the so-called variant quantifier-elimination (VQE) algorithm~\cite{DBLP:journals/jsc/HongD12}, which features singly-exponential complexity in the number of variables. This however requires a careful inspection of whether our problem meets the geometric conditions imposed by VQE; (ii) incorporate relaxation schemes, e.g., Lagrangian relaxation and sum-of-squares decompositions~\cite{DBLP:journals/mp/Parrilo03}, and complement with QE only when the relaxation fails to produce desired results.
\begin{xlrbox}{mybox}
	\begin{tikzpicture}
	\draw (0, 0) node[inner sep=0] (image) {\includegraphics[width=1\linewidth]{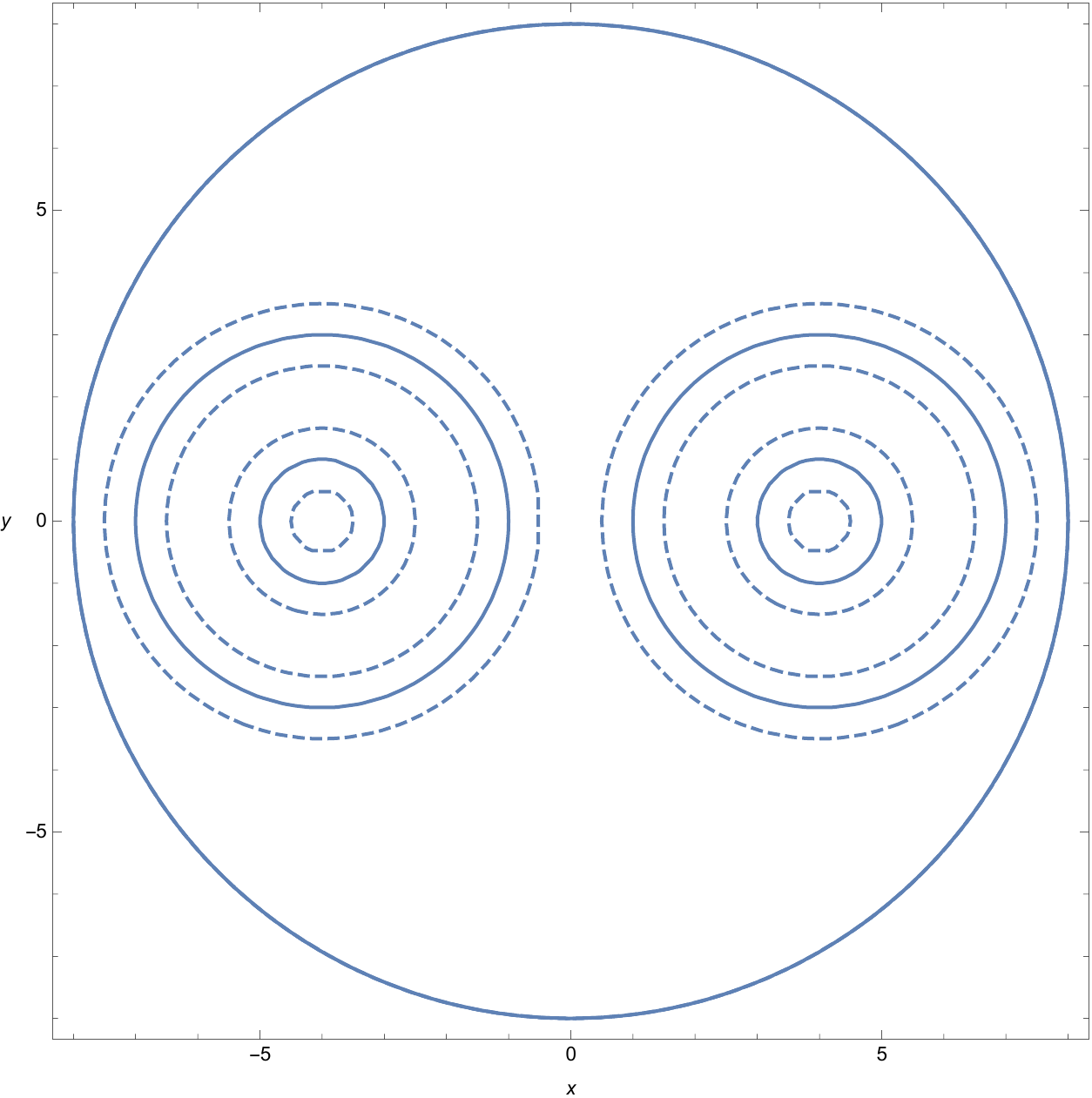}};
	
	\draw[<->] (-3.97,1.77) to +(-0.27cm,0.27cm);
	\node at (-3.97,2.02) {\Large $\epsilon$};
	\draw[latex-latex] (-3.72,1.52) to +(-0.27cm,0.27cm);
	\node at (-3.72,1.77) {\Large $\epsilon$};
	
	\draw[latex-latex] (-2,0.77) to +(0.27cm,0.27cm);
	\node at (-2,1.02) {\Large $\epsilon$};
	\draw[latex-latex] (-1.99,0.78) to +(-0.27cm,-0.27cm);
	\node at (-2.25,0.75) {\Large $\epsilon$};
	
	\draw[latex-latex] (4.52,1.77) to +(0.27cm,0.27cm);
	\node at (4.52,2.02) {\Large $\epsilon$};
	\draw[latex-latex] (4.27,1.52) to +(0.27cm,0.27cm);
	\node at (4.27,1.77) {\Large $\epsilon$};
	
	\draw[latex-latex] (2.55,0.77) to +(-0.27cm,0.27cm);
	\node at (2.55,1.02) {\Large $\epsilon$};
	\draw[latex-latex] (2.54,0.78) to +(0.27cm,-0.27cm);
	\node at (2.81,0.75) {\Large $\epsilon$};
	\end{tikzpicture}
\end{xlrbox}
\begin{figure}[t]
	\centering
	\begin{tabular}{cc}
		\subfloat[$\epsilon$-perturbations in the radii]{~~~~\scalebox{0.345}{\themybox}~~~~\label{fig:face-epsilon-sketch}} &
		%$\quad$
		\subfloat[Interpolant resilient to perturbations]{~~~~\includegraphics[scale=0.33]{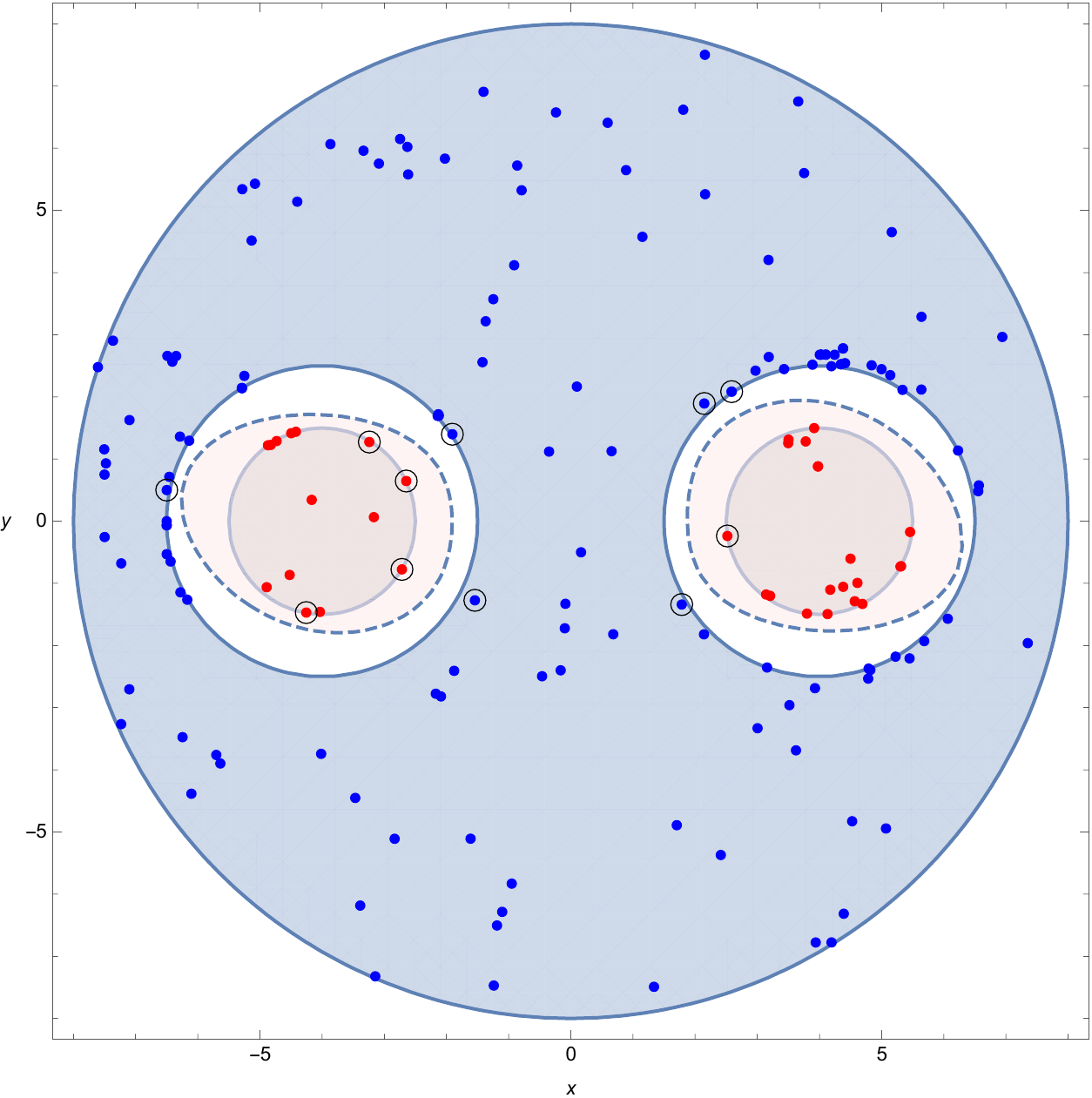}~~~~\label{fig:face-epsilon}}
	\end{tabular}
	\caption{$\epsilon$-Face: introducing perturbations (with $\epsilon$ up to $0.5$) in the Face example. The synthesized interpolant is resilient to any $\epsilon$-perturbation in the radii satisfying $-0.5 \le \epsilon \le 0.5$.}\label{fig:perturbation}
\end{figure}

\vspace*{-.05in}
\section{Taming Perturbations in Parameters.}\label{subsec_robustness}
\vspace*{-.05in}

An interpolant synthesized by the SVM-based technique features inherent robustness due to the way optimal-margin classifier is defined (Sect.~\ref{sec_preliminaries}). That is, the validity of such an interpolant is not easily perturbed by changes (in the feature space) in the formulae to be interpolated. It is straightforward in NIL to deal with interpolation problems under explicitly specified perturbations, which are treated as constraints over fresh variables. An example named $\epsilon$-Face is depicted in Fig.~\ref{fig:perturbation}, which perturbs $\langle \phi, \psi \rangle$ in the Face example as $\phi \define -0.5 \le \epsilon_1 \le 0.5 \wedge ((x+4)^2 + y^2 - (1+\epsilon_1)^2 \le 0 \vee (x-4)^2 + y^2 - (1+\epsilon_1)^2 \le 0)$ and $\psi \define -0.5 \le \epsilon_2 \le 0.5 \wedge x^2 + y^2 - 64 \le 0 \wedge (x+4)^2 + y^2 - (3+\epsilon_2)^2 \ge 0 \wedge (x-4)^2 + y^2 - (3+\epsilon_2)^2 \ge 0$. The synthesized interpolant over common variables of $\phi$ and $\psi$ is 
$\frac{x^4}{139}+\frac{x^3 y}{268}+x^2 \left(\frac{y^2}{39}-\frac{11}{36}\right)+x
\left(-\frac{y^3}{52}-\frac{y^2}{157}-\frac{y}{52}-\frac{1}{116}\right)+\frac{y^4}{25}-
\frac{y^3}{182}+\frac{2 y^2}{19}-\frac{y}{218}+1<0$
which is hence resilient to any $\epsilon$-perturbation in the radii satisfying $-0.5 \le \epsilon \le 0.5$, as illustrated in Fig.~\subref*{fig:face-epsilon}.

\section{Conclusions}\label{sec_conclusion}
\vspace*{-.05in}

We have presented a unified, counterexample-guided method named NIL for generating polynomial interpolants over the general quantifier-free theory of nonlinear arithmetic. Our method is based on classification techniques with space transformations and kernel tricks as established in the community of machine-learning. We proved the soundness of NIL and proposed sufficient conditions for its completeness and convergence. The applicability and effectiveness of our technique are demonstrated experimentally on a collection of representative benchmarks from the literature, including those extracted from program verification. Experimental results indicated that our method suffices to address more interpolation tasks, including those with perturbations in parameters, and in many cases synthesizes simpler interpolants compared with existing approaches.

For future work, we would like to improve the efficiency of NIL by substituting the general purpose quantifier-elimination procedure with alternative methods previously discussed in Sect.~\ref{sec_experiments}. An extension of our approach to cater for the combination of nonlinear arithmetic with EUFs, by resorting to predicate-abstraction techniques~\cite{JhalaPR18}, will be of particular interest. Additionally, we plan to investigate the performance of NIL over different classification techniques, e.g., the widespread regression-based methods~\cite{HastieTF09}, though SVMs are expected to be more competent concerning the robustness and predictability, as also observed in~\cite{Sharma12}.

%\paragraph*{Acknowledgements.}

%\newpage

\bibliographystyle{abbrv}
\bibliography{reference}

%\balancecolumns
\newpage
%\appendix

\setcounter{section}{0}

\begin{subappendices}
\renewcommand{\thesection}{\Alph{section}} % or \arabic{section}
	
\section{Proofs of Theorems}\label{appendix_proofs}

\begin{proof}[of Theorem~\ref{theorem_sound}]
	The argument follows immediately from the checking, done in line~\ref{alg_nil_checking}, of the definition of interpolant (Def.~\ref{def_interpolant}), and the fact that any candidate interpolant
	is of maximum degree $m$.
	\oomit{For the ``if'' counterpart, let $I \in \RR[\xx]_m$ be an interpolant of $\phi$ and $\psi$, then the algorithm will obviously not abort at
	line~\ref{alg_nil_abort1}. Furthermore, $X^+$ and $X^-$ are linearly
	separable in the $\left(\tbinom{m+n}{n}-1\right)$-dimensional
	feature space, where $n$ is the number of common variables of $\phi$
	and $\psi$. Thus by Lemma~\ref{lemma_SVMs_corr}, SVM produces a
	half-space $h$ such that $\forall \vec{x} \in X^+.\ h(\vec{x})>0$
	and $\forall \vec{x} \in X^-.\ h(\vec{x})<0$, and therefore the
	algorithm will not abort at line~\ref{alg_nil_abort2}. Finally, $I$ being an interpolant implies that $\phi \models I$ and $I \wedge \psi \models \bot$ hold and hence, the \textbf{return} statement at line~\ref{alg_nil_return} is reachable and NIL($\phi$,$\psi$,$m$) terminates.} \qed
\end{proof}

% Bohua's version
%\begin{theorem}[Soundness of NIL]
%  NIL($\phi$,$\psi$,$m$) aborts only if there does not exist an
%  interpolant in $\RR[\xx]_m$ between $\phi$ and $\psi$. It
%  terminates and returns $I$ only if $I$ is an interpolant (of maximum
%  degree $m$) between $\phi$ and $\psi$.
%\label{theorem_sound}
%\end{theorem}
%\begin{proof}
%  The second part follows immediately from the checking, done in line
%  \ref{alg_nil_checking}, of the definition of interpolant (Definition
%  \ref{def_interpolant}), and the fact that any candidate interpolant
%  has maximum degree $m$. For the first part, suppose there exists an
%  interpolant of maximum degree $m$ between $\phi$ and $\psi$, then
%  the algorithm will obviously not abort at
%  line~\ref{alg_nil_abort1}. Furthermore, $X^+$ and $X^-$ are linearly
%  separable in the $\left(\tbinom{n}{m+n}-1\right)$-dimensional
%  feature space, where $n$ is the number of common variables of $\phi$
%  and $\psi$. Thus by Lemma~\ref{lemma_SVMs_corr}, SVM produces a
%  half-space $h$ such that $\forall \vec{x} \in X^+.\ h(\vec{x})>0$
%  and $\forall \vec{x} \in X^-.\ h(\vec{x})<0$, and therefore the
%  algorithm will not abort at line~\ref{alg_nil_abort2}. \qed
%\end{proof}

\begin{proof}[of Theorem~\ref{theorem_complete}]
	Suppose the algorithm does not terminate. Then there must be an
	infinite sequence of counterexamples. In particular, the sequence of
	counterexamples added to either $X^+$ or $X^-$ must be
	infinite. Without loss of generality, we assume an infinite sequence
	of counterexamples $\vec{x}_1,\vec{x}_2,\dots$ are added to $X^+$,
	whose initial set of points is $X^+_0$.
	
	Let $\gamma_i$ be the functional margin in $\mathbb{R}^{\tilde{d}}$ for
	the separating hyperplane $P_i$ before the counterexample $\vec{x}_i$ is
	found. We claim that $\gamma_i\ge \gamma$. This is because the convex hull of
	$X^+$ and $X^-$ are subsets of $\llbracket\phi\rrbracket$ and
	$\llbracket\psi\rrbracket$, respectively. Since there exists a
	hyperplane separating $\llbracket\phi\rrbracket$ and
	$\llbracket\psi\rrbracket$ with functional margin $\gamma$, the same
	hyperplane also separates $\conv{X^+}$ and $\conv{X^-}$. Since SVM
	maximizes the functional margin, the actual hyperplane found by SVM
	must have functional margin at least $\gamma$.
	
	Next, we claim $\dist{\vec{x}_i,\vec{x}_j}\ge \gamma/2$ for any $i<j$. Since $\vec{x}_j$ is
	a counterexample found when $X^+$ contains $\vec{x}_1,\dots,\vec{x}_{j-1}$, it
	lies on the other side of the hyperplane $P_i$ from
	$\vec{x}_1,\dots,\vec{x}_{j-1}$. Since the distance between any point on $P_i$
	and the convex hull of $X^+_0\cup\{\vec{x}_1,\dots,\vec{x}_{j-1}\}$ is at least
	$\gamma_i/2\ge \gamma/2$, the same holds for $\vec{x}_j$. In particular, the
	distance between $\vec{x}_j$ and each of the points $\vec{x}_1,\dots,\vec{x}_{j-1}$ is
	at least $\gamma/2$.
	
	In other words, the balls $\mathcal{B}(\vec{x}_i,\gamma/4)$ are disjoint from each
	other. Their total volume is infinite. However, they are contained
	in the set of points whose distance to $\llbracket\phi\rrbracket$ is
	at most $\gamma/4$. This contradicts with the assumption that
	$\llbracket\phi\rrbracket$ is bounded. \qed
\end{proof}

\begin{proof}[of Theorem~\ref{theorem_complete2}]
The proof is analogous to that of Theorem
\ref{theorem_complete}. Again, we assume an infinite sequence of
counterexamples $\vec{x}_1,\vec{x}_2,\dots$ added to $X^+$. Since the
entailment checking is performed with tolerance $\delta$, each
counterexample must have a distance of at least $\delta$ away from
the separating hyperplane. This means that the distance between a
counterexample $\vec{x}_n$ and each existing one
$\vec{x}_1,\dots,\vec{x}_{n-1}$ is at least $\delta$. Hence, the
balls $\mathcal{B}(\vec{x}_i,\delta/2)$ are disjoint from each
other and their total volume is infinite. However, they are
contained in the set of points whose distance to
$\llbracket\phi\rrbracket$ is at most $\delta/2$, which contradicts
the fact that $\llbracket\phi\rrbracket$ is bounded. \qed
\end{proof}

\begin{proof}[of Theorem~\ref{theorem_convergence}]
Each iteration of NIL$^*_{\delta,B}(\phi,\psi,m)$ can be considered
as a run of algorithm NIL$_\delta$ with tolerance $\delta=\delta_i$,
where the two regions are replaced by
$\llbracket\phi\rrbracket\cap[-B,B]^{\tilde{d}}$ and
$\llbracket\psi\rrbracket\cap[-B,B]^{\tilde{d}}$. The two regions are
clearly bounded, hence by Theorem \ref{theorem_complete2}, each
iteration of NIL$^*_{\delta,B}(\phi,\psi,m)$ terminates. Given any
point $p$ in the interior of either $\llbracket \phi \rrbracket$ or
$\llbracket \psi \rrbracket$, we need to show that there exists some
integer $K_p$ such that $I_k$ classifies $p$ correctly for all
$k\ge K_p$. Since each iteration of the algorithm terminates, it
suffices to show the existence of an integer $i$ such that $p$ is
classified correctly after the $i$-th iteration. We take $i$
sufficiently large so that the open ball $\mathcal{B}(p,\delta_i)$
lies entirely in the region containing $p$ (which is possible since
$p$ is in the interior and $\delta_i$ converges to 0), and the
absolute value of $p$ in each dimension is at most $B_i$ (which is
possible since $B_i$ tends to infinity). It is then clear that $p$
will be correctly classified after the $i$-th iteration. \qed
\end{proof}

\end{subappendices}

% That's all folks!
\end{document}